\pdfoutput=1
\documentclass[fleqn, usenatbib, usedcolumn]{mnras}

\newif\iflocal
\localfalse

\usepackage{graphicx}
\usepackage{xspace}
\usepackage{amsmath}
\usepackage{framed} 
\usepackage{txfonts}
\usepackage{epstopdf}
\usepackage{color}
\usepackage{rotating}
\usepackage{natbib}
\usepackage{ulem}
\usepackage{xspace}
\usepackage{sistyle}

\iflocal
\def\includedir{/Users/benedito/University/docs/latex}
\def\figdir{figs}
\else
\def\includedir{.}
\def\figdir{.}
\fi

\newcommand{\figmn}[1]{Fig.~\ref{#1}\xspace}
\newcommand{\eqmn}[1]{equation~(\ref{#1})\xspace}
\newcommand{\eqmnb}[1]{equation~\ref{#1}\xspace}

\def\gtsima{$\; \buildrel > \over \sim \;$}
\def\ltsima{$\; \buildrel < \over \sim \;$}
\def\prosima{$\; \buildrel \propto \over \sim \;$}
\def\gsim{\lower.7ex\hbox{\gtsima}}
\def\lsim{\lower.7ex\hbox{\ltsima}}
\def\simgt{\lower.7ex\hbox{\gtsima}}
\def\simlt{\lower.7ex\hbox{\ltsima}}
\def\simpr{\lower.7ex\hbox{\prosima}}

\newcommand{\dnorm}[2]{\frac{{\rm d} #1}{{\rm d} #2}}
\newcommand{\dnorminl}[2]{{\rm d} #1 / {\rm d} #2}

\newcommand{\vect}[1]{{\pmb #1}}

\newcommand{\dotp}{\boldsymbol{\cdot}}

\newcommand{\kpch}{{h^{-1}{\rm kpc}}}
\newcommand{\mpch}{h^{-1}{\rm {Mpc}}}

\newcommand{\gyr}{{\rm Gyr}}

\newcommand{\msun}{{M_{\odot}}}
\newcommand{\msunh}{h^{-1} M_\odot}

\newcommand{\tdyn}{t_{\rm dyn}}

\newcommand{\LCDM}{$\Lambda$CDM\xspace}
\newcommand{\rhoc}{\rho_{\rm c}}
\newcommand{\rhom}{\rho_{\rm m}}

\def\sparta{\textsc{Sparta}\xspace}
\def\moria{\textsc{Moria}\xspace}

\def\colossus{\textsc{Colossus}\xspace}

\def\rockstar{\textsc{Rockstar}\xspace}
\def\consistenttrees{\textsc{Consistent-Trees}\xspace}

\def\planck{Planck\xspace}
\def\wmap{WMAP7\xspace}

\def\erebos{Erebos\xspace}

\def\rmm{{\rm m}}

\def\rmt{{\rm t}}

\def\rmL{{\rm L}}

\def\rmR{{\rm R}}

\def\deltac{\delta_{\rm c}}
\def\rs{r_{\rm s}}
\def\vmax{V_{\rm max}}

\def\vr{v_{\rm r}}

\def\gammadyn{\Gamma_{\rm dyn}}

\def\neff{n_{\rm eff}}

\def\rhoorb{\rho_{\rm orb}}
\def\rhoinf{\rho_{\rm inf}}
\def\rhotot{\rho_{\rm tot}}

\def\rvir{R_{\rm vir}}

\def\mtom{M_{\rm 200m}}
\def\rtom{R_{\rm 200m}}
\def\ctom{c_{\rm 200m}}
\def\vtom{V_{\rm 200m}}
\def\ntom{N_{\rm 200m}}
\def\nutom{\nu_{\rm 200m}}

\def\mtoc{M_{\rm 200c}}
\def\rtoc{R_{\rm 200c}}

\def\rfoc{R_{\rm 500c}}

\def\rtombnd{R_{\rm 200m,bnd}}
\def\rtomall{R_{\rm 200m,all}}
\def\mtombnd{M_{\rm 200m,bnd}}
\def\mtomall{M_{\rm 200m,all}}
\def\nutomall{\nu_{\rm 200m,all}}

\def\rsp{R_{\rm sp}}

\newcommand{\rt}{r_\rmt}

\usepackage{etoolbox}
\makeatletter
\patchcmd{\NAT@citex}
  {\@citea\NAT@hyper@{\NAT@nmfmt{\NAT@nm}\NAT@date}}
  {\@citea\NAT@nmfmt{\NAT@nm}\NAT@hyper@{\NAT@date}}
  {}% Do nothing if patch works
  {}% Do nothing if patch fails
\patchcmd{\NAT@citex}
  {\@citea\NAT@hyper@{%
     \NAT@nmfmt{\NAT@nm}%
     \hyper@natlinkbreak{\NAT@aysep\NAT@spacechar}{\@citeb\@extra@b@citeb}%
     \NAT@date}}
  {\@citea\NAT@nmfmt{\NAT@nm}%
   \NAT@aysep\NAT@spacechar%
   \NAT@hyper@{\NAT@date}}
  {}% Do nothing if patch works
  {}% Do nothing if patch fails
\patchcmd{\NAT@citex}
  {\@citea\NAT@hyper@{%
     \NAT@nmfmt{\NAT@nm}%
     \hyper@natlinkbreak{\NAT@spacechar\NAT@@open\if*#1*\else#1\NAT@spacechar\fi}%
       {\@citeb\@extra@b@citeb}%
     \NAT@date}}
  {\@citea\NAT@nmfmt{\NAT@nm}%
   \NAT@spacechar\NAT@@open\if*#1*\else#1\NAT@spacechar\fi%
   \NAT@hyper@{\NAT@date}}
  {}% Do nothing if patch works
  {}% Do nothing if patch fails
\makeatother

\iflocal
\def\figdir{figs}
\else
\def\figdir{.}
\fi

\defcitealias{diemer_13_scalingrel}{DKM13}
\defcitealias{diemer_14}{DK14}
\defcitealias{diemer_15}{DK15}
\defcitealias{diemer_17_sparta}{D17}
\defcitealias{diemer_20_catalogs}{D20}

\newcommand{\papertwo}{Paper II\xspace}
\newcommand{\paperthree}{Paper III\xspace}
\newcommand{\papertwothree}{Papers II and III\xspace}

\title[Dynamics-based halo density profiles I]{A dynamics-based density profile for dark haloes -- I. Algorithm and basic results}

\author[Diemer]{Benedikt Diemer\thanks{Email: \href{mailto:diemer@umd.edu}{diemer@umd.edu}}
\vspace{1mm}
\\
Department of Astronomy, University of Maryland, College Park, MD 20742, USA \\
}

\date{Accepted: 2022 March 28; Revised: 2022 March 23; Received: 2021 December 8}
\pubyear{2022}

\begin{document}
\label{firstpage}
\pagerange{\pageref{firstpage}--\pageref{lastpage}}
\maketitle

%%%%%%%%%%%%%%%%%%%%%%%%%%%%%%%%%%%%%%%%%%%%%%%%%%%
% ABSTRACT
%%%%%%%%%%%%%%%%%%%%%%%%%%%%%%%%%%%%%%%%%%%%%%%%%%%

\begin{abstract}
The density profiles of dark matter haloes can potentially probe dynamics, fundamental physics, and cosmology, but some of the most promising signals reside near or beyond the virial radius. While these scales have recently become observable, the profiles at large radii are still poorly understood theoretically, chiefly because the distribution of orbiting matter (the one-halo term) is partially concealed by particles falling into halos for the first time. We present an algorithm to dynamically disentangle the orbiting and infalling contributions by counting the pericentric passages of billions of simulation particles. We analyse dynamically split profiles out to $10\ \rtom$ across a wide range of halo mass, redshift, and cosmology. We show that the orbiting term experiences a sharp truncation at the edge of the orbit distribution. Its sharpness and position are mostly determined by the mass accretion rate, confirming that the entire profile shape primarily depends on halo dynamics and secondarily on mass, redshift, and cosmology. The infalling term also depends on the accretion rate for fast-accreting haloes but is mostly set by the environment for slowly accreting haloes, leading to a diverse array of shapes that does not conform to simple theoretical models. While the resulting scatter in the infalling term reaches $1$ dex, the scatter in the orbiting term is only between $0.1$ and $0.4$ dex and almost independent of radius. We demonstrate a tight correspondence between the redshift evolution in \LCDM and the slope of the matter power spectrum. Our code and data are publicly available.
\end{abstract}


%%%%%%%%%%%%%%%%%%%%%%%%%%%%%%%%%%%%%%%%%%%%%%%%%%%
% INTRODUCTION
%%%%%%%%%%%%%%%%%%%%%%%%%%%%%%%%%%%%%%%%%%%%%%%%%%%

\section{Introduction}
\label{sec:intro}

Although dark matter accounts for the majority of mass in the Universe, it is notoriously hard to observe. The most promising place to at least infer some of its properties are dark matter haloes (hereafter simply haloes) because their roughly spherical nature makes them more orderly than filaments, walls, or (arguably) voids, and because the highest densities are reached at halo centres. However, those centres are also strongly affected by baryonic physics that we cannot claim to fully model at this point. Thus, it is critical to understand haloes in detail out to large radii, where they join into the surrounding large-scale structure.

The shapes of haloes are complex and not spherically symmetric, especially at their interface with the cosmic web \citep[e.g.,][]{bond_96_filaments, jing_02, allgood_06, mansfield_17}. However, the full three-dimensional density structure of haloes is challenging to model theoretically \citep[e.g.,][]{vogelsberger_09, vogelsberger_11_similarity}, complex to quantify from simulations \citep{lilley_18_expansions}, and difficult to observe \citep{shin_18}. Instead, spherically averaged density profiles (hereafter abbreviated as `profiles') provide a simple way to describe halo structure. This paper continues a long history of efforts to theoretically understand density profiles, be it via spherical collapse models \citep[e.g.,][]{gunn_72}, from $N$-body simulations \citep[e.g.,][]{frenk_85, frenk_88, barnes_87,  dubinski_91}, or with empirical fitting functions such as those of \citet{einasto_65}, \citet{hernquist_90}, and \citet[][hereafter NFW]{navarro_97}. 

This enormous body of work has established a number of basic facts. The inner profile, roughly within $\rtom$ (the radius enclosing an average overdensity of $200$ times the cosmic mean), is dominated by particles orbiting in the halo \citep[e.g.,][]{cuesta_08, diemand_08}. Although often labelled the `one-halo term', we call this component the `orbiting term'. It gradually steepens with radius, a generic consequence of hierarchical collapse from initial peaks with a range of shapes \citep[e.g.,][]{huss_99, dalal_10, ludlow_13}. At large radii (roughly beyond $\rtom$), the profiles become dominated by matter that is falling into the halo for the first time and, at even larger radii, by the statistical contribution from other haloes. These two components are often collectively mislabelled as the `two-halo term'. We will refer to them as the `infalling term' because that contribution dominates out to the radii we consider ($10\ \rtom$).

Despite this progress, two key questions remain. First, how is dark matter distributed near the halo centre, where baryons dominate? Much work has been focused on this question \citep{moore_99_collapse, jing_00_profiles1, navarro_04, navarro_10, kazantzidis_06}, but its solution demands an accurate understanding of complex baryonic physics \citep[e.g.,][]{pontzen_12, dicintio_14, velliscig_14, schaller_15, wang_20_profiles}. The second question is the focus of this paper: what is the structure near the transition between the orbiting and infalling terms, and beyond the halo edge? While feedback can eject gas to those radii, baryonic effects are much weaker near the transition region \citep{schneider_19, arico_20, sorini_22} and do not fundamentally alter the features we are interested in \citep{lau_15, oneil_21}.

The halo outskirts ($r \gsim 0.5\ \rtom$) have recently received renewed attention because they turn out to be sensitive to otherwise inaccessible halo properties such as the mass accretion rate (\citealt{diemer_14}, hereafter \citetalias{diemer_14}; see also \citealt{xhakaj_20, xhakaj_21}), the nature of dark matter \citep{banerjee_20}, and deviations from General Relativity \citep{adhikari_18_mog, contigiani_19_symmetron}. At the same time, rapid observational progress has led to high-accuracy constraints based on profiles of satellites in individual clusters \citep{rines_13, tully_15, patej_16}, in stacked clusters \citep{more_16, baxter_17, nishizawa_18, zuercher_19, murata_20}, and in weak-lensing profiles \citep{mandelbaum_06_group_profiles, umetsu_11, umetsu_17, chang_18, contigiani_19_wl, shin_19_rsp, shin_21}. Deviations from simple fitting functions for the orbiting  profile have long been known to bias weak lensing masses \citep{becker_11, oguri_11}, but the recently acquired high signal-to-noise ratio in the outer profiles has made detailed theoretical modelling even more pressing.
 
 \begin{figure}
\centering
\includegraphics[trim =  5mm 3mm 4mm 4mm, clip, width=0.46\textwidth]{\figdir/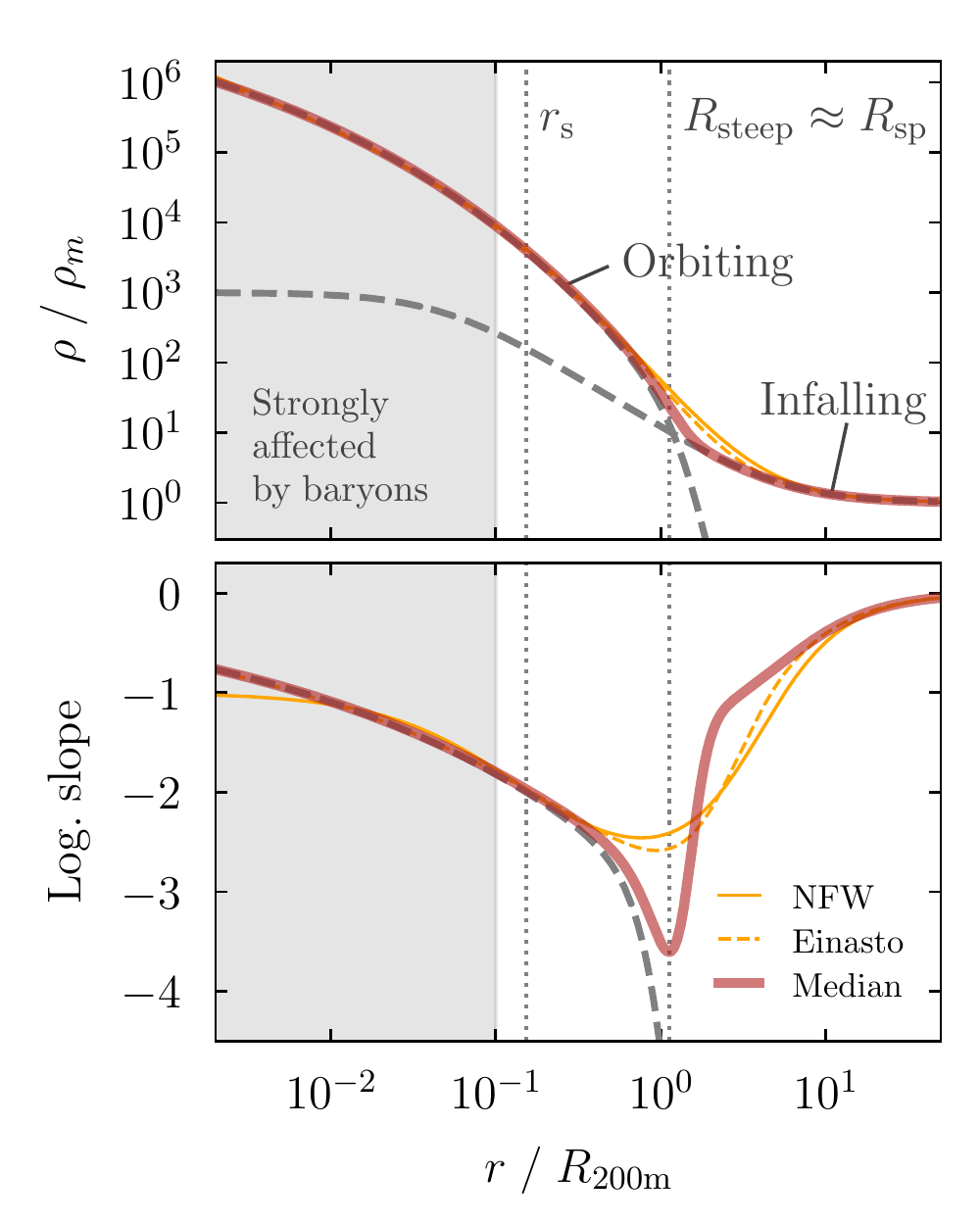}
\caption{Schematic depiction of the orbiting and infalling contributions to halo density profiles (dashed gray lines). In general, we can measure only the total profile (red), which makes it difficult to discern the true shape of the orbiting term. The gray shading marks $r < 0.1\ \rtom$, roughly $10$ times the size of galaxies \citep{kravtsov_13}, although baryons play an important role even at larger radii. The bottom panel shows the logarithmic slope, $\dnorminl{\ln \rho}{\ln r}$. For this example, we consider the typical (median) profile of a cluster halo with $\mtom = 5 \times 10^{14} \msunh$, which corresponds to a median concentration of $\ctom \approx 6.5$ \citep{diemer_19_cm}. The gray/red profiles show the fitting function of \citetalias{diemer_14}, which accurately fits the total profiles, but the behaviour of the orbiting term near its cut-off is poorly constrained. This edge represents the extent of the largest particle orbits and is also known as the splashback radius, $\rsp$. This boundary is often operationally defined as the location where the logarithmic slope is steepest, $R_{\rm steep}$, but this radius is the result of a complicated interplay between the infalling and orbiting components. The overall shape of the profiles in the transition regions is not well matched by classical fitting functions such as the Einasto or NFW profiles (yellow).}
\label{fig:comic}
\end{figure}

The paramount difficulty in modelling the outer profile arises from the superposition of the orbiting and infalling terms, as illustrated in \figmn{fig:comic}. These terms (gray lines) trade off near the steepest part of the total profile (red). By construction, the steeply decreasing orbiting term leaves the infalling term to dominate at large radii, meaning that the true shape of the orbiting term near its truncation cannot be measured. In the figure, it is approximated by the \citetalias{diemer_14} fitting function, but its asymptotic functional form is essentially unconstrained. The shape near the truncation is particularly important because the orbiting profile contains valuable information about particle dynamics. For example, the edge of the apocentre distribution, also known as the splashback radius, marks a physically meaningful halo boundary (\citetalias{diemer_14}, \citealt{adhikari_14}, \citealt{more_15}). Approximations such as the Einasto or NFW profiles were designed to fit at $r \lsim \rtom$ and thus do not match the behaviour in the transition region (Fig.~\ref{fig:comic}). Additional theoretical modelling and fitting functions have helped to elucidate the nature of the infalling term \citep{betancortrijo_06, prada_06, tavio_08, oguri_11}, but they could not quite fit the sharp drop near the transition region \citepalias{diemer_14}. 

\begin{figure*}
\centering
\includegraphics[trim =  3mm 10mm 3mm 2mm, clip, width=\textwidth]{\figdir/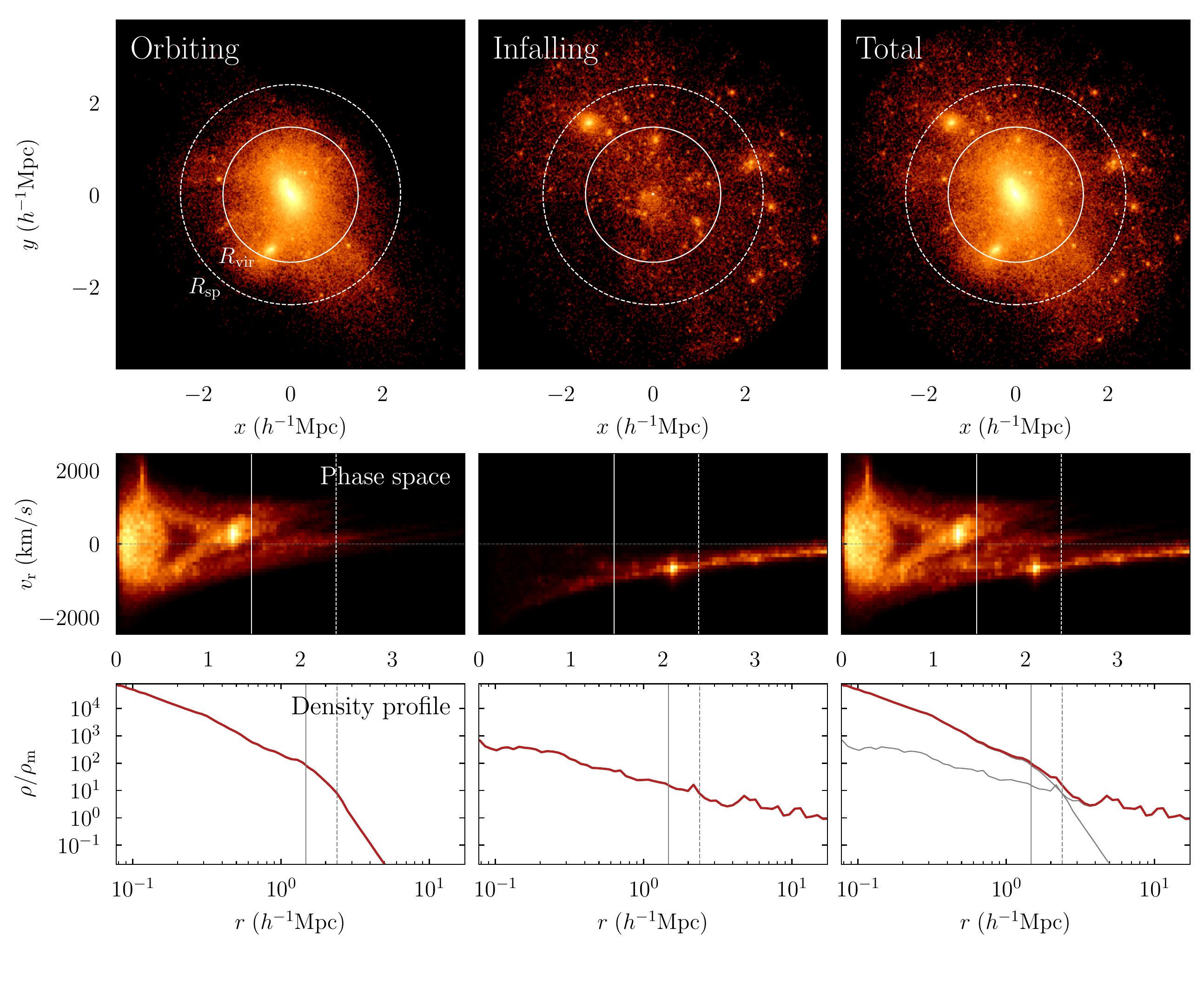}
\caption{The dynamics and density structure of a halo with a relatively low accretion rate ($\Gamma = 1.3$, from TestSim200 at $z \approx 0$). The halo's particles are split by our orbit counting algorithm into orbiting particles (which have had at least one pericentre, left column) and infalling particles (centre column). The top panels show the projected density field using an SPH-like smoothing kernel with a fixed size and a logarithmic color scale. The second row shows the corresponding phase-space density structure (on a linear color scale). The bottom row shows the corresponding density profiles. For comparison, we show $\rvir$ (solid lines) and $R_{\rm sp,90\%}$ (dashed lines) as measured by \sparta. For this halo, the infalling and orbiting terms are separated relatively clearly in phase space.}
\label{fig:viz1}
\end{figure*}

\begin{figure*}
\centering
\includegraphics[trim =  3mm 10mm 3mm 2mm, clip, width=\textwidth]{\figdir/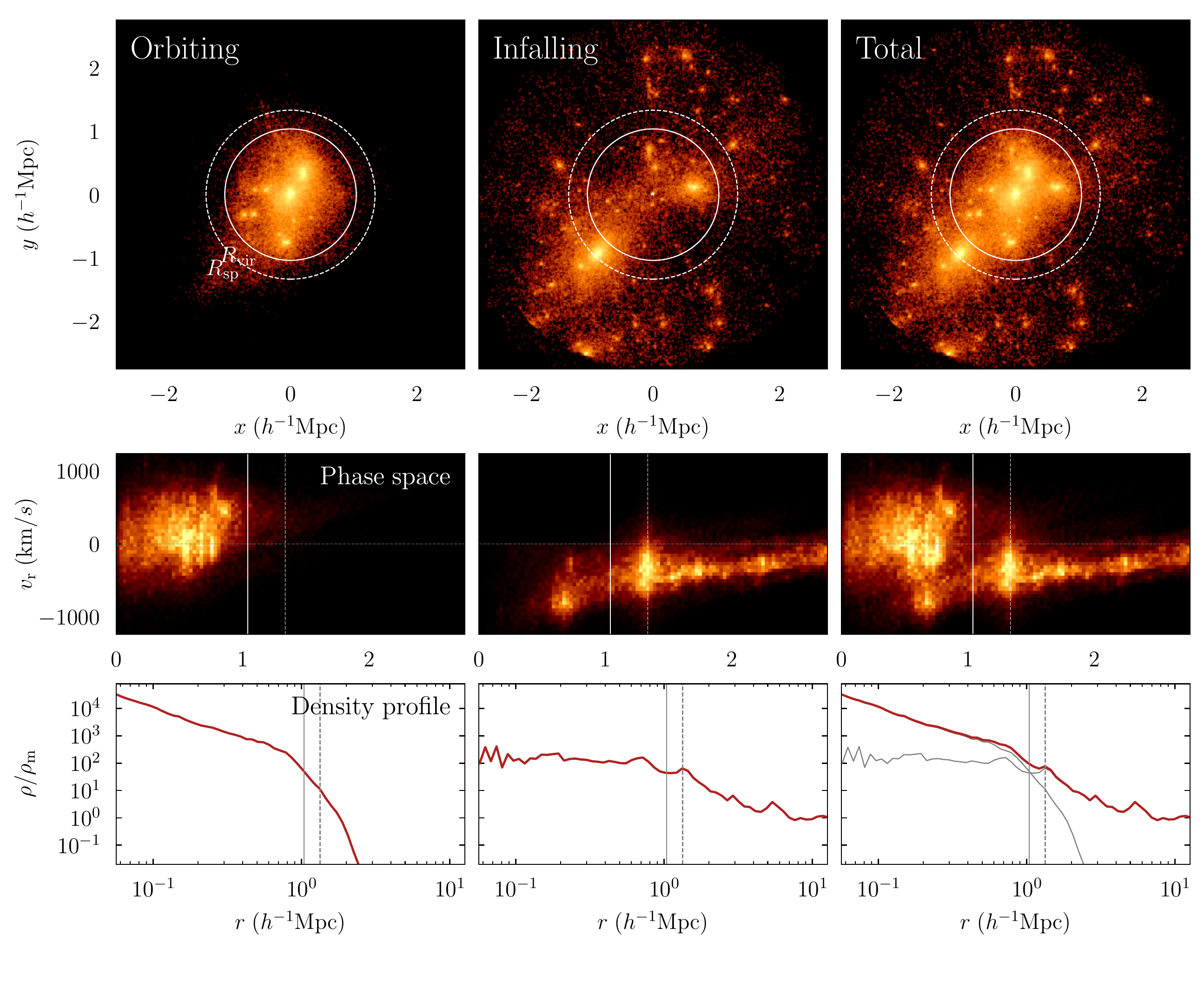}
\caption{Same as \figmn{fig:viz1} but for a rapidly accreting halo with $\Gamma = 3$. The phase-space structure of the orbiting particles is somewhat irregular. Moreover, the infalling term contains two massive and many smaller subhaloes. The velocity dispersion of the most massive subhalo (at a radius of about $1.3 \mpch$) is large enough to give some of its particles a positive total $\vr$, highlighting why our algorithm cannot solely rely on the radial velocity as an indicator of pericentres. }
\label{fig:viz2}
\end{figure*}

The key insight is that the profiles are intrinsically dynamical in nature. It has long been known that the orbiting profiles depend on the growth history of haloes \citep{navarro_96, bullock_01, wechsler_02, zhao_03_mah, tasitsiomi_04_clusterprof, kazantzidis_06, ludlow_13}. Moreover, \citetalias{diemer_14} showed that the profile shape at all radii depends on the current mass accretion rate. While statistical mechanics has previously been the basis of theoretical models for the orbiting term \citep{king_66, hjorth_10_darkexp1, pontzen_13}, the dependence on growth rate goes beyond the velocity dispersion of particles and highlights that a full understanding of the transition region must depend on a dynamical distinction between infalling and orbiting particles. A cut in phase space cannot achieve this goal because the two populations overlap in the radial range of interest \citep[see Figs.~\ref{fig:viz1} and \ref{fig:viz2};][]{fukushige_01, cuesta_08, diemand_08, sugiura_20}, although approximate cuts are useful when classifying observed satellite distributions \citep[e.g.,][]{oman_13, tomooka_20, aung_21_phasespace}.

In this paper, we present a novel, reliable algorithm to systematically perform a dynamical distinction for billions of particles. Specifically, we count the pericentric passages for each halo particle in large $N$-body simulations and construct separate orbiting and infalling profiles (see \citealt{bakels_21} for a similar algorithm applied to subhaloes and \citealt{sugiura_20} for an apocentre-based scheme). Our algorithm is implemented in the publicly available \sparta framework, which is described in \citet[][hereafter \citetalias{diemer_17_sparta}]{diemer_17_sparta} and \citet[][hereafter \citetalias{diemer_20_catalogs}]{diemer_20_catalogs}.

The paper is organized as follows. In Section~\ref{sec:methods}, we describe our simulations and the dynamical splitting algorithm. Additional convergence tests are shown in Appendix~\ref{sec:tests}. In Section~\ref{sec:results}, we present the most important insights into the separate orbiting and infalling profiles, referring the reader to online figures for a more complete accounting of the numerical data (see \href{http://www.benediktdiemer.com/data/}{benediktdiemer.com/data}). We discuss theoretical questions in Section~\ref{sec:discussion} and summarize our findings in Section~\ref{sec:conclusion}. This paper is the first of a series. The second paper (hereafter \papertwo) will present improved fitting functions for the orbiting and infalling profiles. \paperthree will investigate the parameter space of these functions and connect it to halo properties. Throughout the paper series, we follow the notation of \citetalias{diemer_20_catalogs} (see also Section~\ref{sec:methods:defs}). The code and profile data used for this paper are publicly available.

%%%%%%%%%%%%%%%%%%%%%%%%%%%%%%%%%%%%%%%%%%%%%%%%%%%
% METHODS & SIMULATIONS
%%%%%%%%%%%%%%%%%%%%%%%%%%%%%%%%%%%%%%%%%%%%%%%%%%%

\vspace{-1.1cm}

\section{Methods and algorithms}
\label{sec:methods}

In this section, we give a detailed account of the numerical methods that underlie our results. We begin by introducing our simulation data in Section~\ref{sec:methods:sims} and by reviewing the definitions of important halo-related quantities in Section~\ref{sec:methods:defs}. We briefly summarize the \sparta code in Section~\ref{sec:methods:sparta} but refer the reader to \citetalias{diemer_17_sparta} and \citetalias{diemer_20_catalogs} for details. Our new algorithms consist of two main parts, namely counting the orbits of particles (Section~\ref{sec:methods:oct}) and constructing density profiles from those particles (Sections~\ref{sec:methods:prf} and \ref{sec:methods:prf_av}). We discuss how we select and combine halo samples in Sections~\ref{sec:methods:halo_sel} and \ref{sec:methods:combined}, and we subject these methods to convergence tests in Appendix~\ref{sec:tests}.

\vspace{-0.5cm}

\subsection{$N$-body simulations and halo finding}
\label{sec:methods:sims}

We use the \erebos suite of dissipationless $N$-body simulations (Table~\ref{table:sims}). This suite contains simulations of different box sizes and resolutions as well as two \LCDM cosmologies and self-similar universes. The first \LCDM cosmology is the same as that of the Bolshoi simulation \citep{klypin_11} and is consistent with \wmap \citep{komatsu_11}, namely, flat \LCDM with $\Omega_{\rm m} = 0.27$, $\Omega_{\rm b} = 0.0469$, $\sigma_8 = 0.82$, and $n_{\rm s} = 0.95$. The second is a \planck-like cosmology with $\Omega_{\rm m} = 0.32$, $\Omega_{\rm b} = 0.0491$, $h = 0.67$, $\sigma_8 = 0.834$, and $n_{\rm s} = 0.9624$ \citep[][]{planck_14}. These two cosmologies span the currently favoured ranges of the parameters that are most important for structure formation, such as $\Omega_{\rm m}$. 

To investigate the universality of profiles, it is important to consider universes with very different power spectra. We rely on four self-similar Einstein-de Sitter universes with power-law initial spectra, $P \propto k^n$, with $n = -1$, $-1.5$, $-2$, and $-2.5$. In these simulations, length and time are scale-free, meaning that the density profiles are independent of redshift when rescaled to meaningful physical units. Their differences isolate the impact of the initial power spectrum \citep[e.g.,][]{efstathiou_88, cole_96, knollmann_08, diemer_15, ludlow_17}.

The power spectra of the \LCDM simulations were generated using \textsc{Camb} \citep{lewis_00}. The power spectra were translated into initial conditions using the \textsc{2LPTic} code \citep{crocce_06}. The simulations were run with \textsc{Gadget2} \citep{springel_05_gadget2}. We identify haloes and subhaloes using the phase-space friends-of-friends \citep{davis_85} halo finder \textsc{Rockstar} \citep{behroozi_13_rockstar} and construct merger trees using the \textsc{Consistent-Trees} code \citep{behroozi_13_trees}. These catalogues are described in detail in \citetalias{diemer_20_catalogs}. The halo centre is taken to be the average position of the most bound particles, a procedure that minimizes the Poisson noise in the position \citep{behroozi_13_rockstar}. The velocity is taken to be the average of all particles within $0.1\ \rtom$.

\subsection{Definitions of halo mass, radius, and accretion rate}
\label{sec:methods:defs}

\begin{table*}
\centering
\caption{The \erebos suite of $N$-body simulations. $L$ denotes the box size in comoving units, $N^3$ the number of particles, $m_{\rm p}$ the particle mass, and $\epsilon$ the force softening length in comoving units. The redshift range of each simulation is determined by the first and last redshifts $z_{\rm initial}$ and $z_{\rm final}$, but snapshots were output only between $z_{\rm f-snap}$ and $z_{\rm final}$. The earliest snapshots of some simulations do not yet contain any haloes, so that the first catalogue with haloes is output at $z_{\rm f-cat}$; the \sparta and \moria data also begin at that redshift. The cosmological parameters are given in Section~\ref{sec:methods:sims}. Here, `PL' indicates self-similar cosmologies with a power-law initial spectrum with slope $n$. The references correspond to \citet[][\citetalias{diemer_13_scalingrel}]{diemer_13_scalingrel}, \citet[][\citetalias{diemer_14}]{diemer_14}, \citet[][\citetalias{diemer_15}]{diemer_15}, and \citet[][\citetalias{diemer_17_sparta}]{diemer_17_sparta}. Our system for choosing force resolutions is discussed in \citetalias{diemer_14}. Some previous papers on the \erebos simulations (including \citetalias{diemer_15}, \citetalias{diemer_17_sparta}, and \citetalias{diemer_20_catalogs}) erroneously claimed $\epsilon$ to be fixed in physical units, but this error had no impact on their calculations.}
\label{table:sims}
\begin{tabular}{lcccccccccccc}
\hline
Name & $L$ & $N^3$ & $m_{\rm p}$ & $\epsilon$ & $\epsilon$ &$z_{\rm initial}$ & $z_{\rm final}$ & $N_{\rm snaps}$ & $z_{\rm f-snap}$ & $z_{\rm f-cat}$ & Cosmology & Reference \\
 & $(\mpch)$ & & $(\msunh)$ & $(\kpch)$ & $(L / N)$ & & & & & & & \\
\hline
L2000-WMAP7 & $2000$ & $1024^3$ & $5.6 \times 10^{11}$  & $65$  & $1/30$ & $49$ & $0$ & $100$ & $20$ & $4.2$ & WMAP7 & \citetalias{diemer_15} \\
L1000-WMAP7 & $1000$ & $1024^3$ & $7.0 \times 10^{10}$ & $33$ & $1/30$ & $49$ & $0$ &  $100$ & $20$ & $6.2$ & WMAP7 & \citetalias{diemer_13_scalingrel} \\
L0500-WMAP7 & $500$  & $1024^3$ & $8.7 \times 10^{9}$  & $14$ & $1/35$  & $49$ & $0$ &  $100$ & $20$ & $8.8$ & WMAP7 & \citetalias{diemer_14} \\
L0250-WMAP7 & $250$  & $1024^3$ & $1.1 \times 10^{9}$  & $5.8$  & $1/42$  & $49$ & $0$ &  $100$ & $20$ & $11.5$ & WMAP7 & \citetalias{diemer_14} \\
L0125-WMAP7 & $125$  & $1024^3$ & $1.4 \times 10^{8}$  & $2.4$  & $1/51$  & $49$ & $0$ &  $100$ & $20$ & $14.5$ & WMAP7 & \citetalias{diemer_14} \\
L0063-WMAP7 & $62.5$ & $1024^3$ & $1.7 \times 10^{7}$  & $1.0$  & $1/60$ & $49$ & $0$ &  $100$ & $20$ & $17.6$ & WMAP7 & \citetalias{diemer_14} \\
L0031-WMAP7 & $31.25$ & $1024^3$ & $2.1 \times 10^{6}$  & $0.25$  & $1/122$ & $49$ & $2$ &  $64$ & $20$ & $20$ & WMAP7 & \citetalias{diemer_15} \\
L0500-Planck & $500$  & $1024^3$ & $1.0 \times 10^{10}$  & $14$ & $1/35$  & $49$ & $0$ &  $100$ & $20$ & $9.1$ & Planck & \citetalias{diemer_15} \\
L0250-Planck & $250$  & $1024^3$ & $1.3 \times 10^{9}$  & $5.8$  & $1/42$  & $49$ & $0$ &  $100$ & $20$ & $12.3$ & Planck & \citetalias{diemer_15} \\
L0125-Planck & $125$  & $1024^3$ & $1.6 \times 10^{8}$  & $2.4$  & $1/51$  & $49$ & $0$ &  $100$ & $20$ & $15.5$ & Planck & \citetalias{diemer_15} \\
L0100-PL-1.0 & $100$  & $1024^3$ & $2.6 \times 10^{8}$  & $0.5$  & $1/195$  & $119$ & $2$ & $64$ & $20$ & $20$ & PL, $n=-1.0$ & \citetalias{diemer_15} \\
L0100-PL-1.5 & $100$  & $1024^3$ & $2.6 \times 10^{8}$  & $0.5$  & $1/195$  & $99$ & $1$ & $78$ & $20$ & $20$ & PL, $n=-1.5$ & \citetalias{diemer_15} \\
L0100-PL-2.0 & $100$  & $1024^3$ & $2.6 \times 10^{8}$  & $1.0$  & $1/98$  & $49$ & $0.5$ & $100$ & $20$ & $15.5$ & PL, $n=-2.0$ & \citetalias{diemer_15} \\
L0100-PL-2.5 & $100$  & $1024^3$ & $2.6 \times 10^{8}$  & $1.0$  & $1/98$  & $49$ & $0$ & $100$ & $20$ & $5.4$ & PL, $n=-2.5$ & \citetalias{diemer_15} \\
TestSim200 & $62.5$  & $256^3$  & $1.1 \times 10^{9}$  & $5.8$  & $1/42$  & $49$ & $-0.1$ & $193$ & $9$  & $9$ & WMAP7 & \citetalias{diemer_17_sparta} \\
TestSim100 & $62.5$  & $256^3$  & $1.1 \times 10^{9}$  & $5.8$  & $1/42$  & $49$ & $-0.1$ & $96$  & $9$  & $9$ & WMAP7 & \citetalias{diemer_17_sparta} \\
TestSim50  & $62.5$  & $256^3$  & $1.1 \times 10^{9}$  & $5.8$  & $1/42$  & $49$ & $-0.1$ & $48$  & $9$  & $9$ & WMAP7 & \citetalias{diemer_17_sparta} \\
\hline
\end{tabular}
\end{table*}

We consider three types of halo boundary definitions: bound-only spherical overdensity (SO) radii, all-particle SO radii, and splashback radii. Given a halo radius $R_{\rm X}$, we denote the mass enclosed within it as $M_{\rm X}$ and the corresponding number of particles as $N_{\rm X}$. 

Bound-only radii are calculated by \rockstar, which removes gravitationally unbound particles from friends-of-friends groups and subgroups in six-dimensional phase space. SO radii are then computed by finding the outermost radius where the enclosed density falls below the given SO threshold. We mostly consider a threshold of $200$ times the mean matter density of the Universe, $\rhom$; we denote the corresponding boundary and enclosed mass as $\rtombnd$ and $\mtombnd$. Alternatively, one can use a threshold based on the critical density of the Universe, $\rhoc$, leading to definitions such as $\rtoc$ and $\mtoc$. Finally, we compute the varying virial overdensity definition, $\rvir$, using the approximation of \citet{bryan_98}.

However, throughout the paper we consider density profiles that include all particles, whether gravitationally bound or not. Thus, we mostly use the corresponding SO radii and masses, particularly $\rtomall$ and $\mtomall$, which we abbreviate to $\rtom$ and $\mtom$. We compute these quantities using \sparta (Section~\ref{sec:methods:sparta}). For the vast majority of haloes, the difference between all-particle and bound-only masses is small, but it can matter for haloes where the density profile contains significant contributions from other, neighbouring haloes. We deal with such cases in Section~\ref{sec:methods:halo_sel}, using the difference between $\mtomall$ and $\mtombnd$ as an indicator of a dense environment. Finally, splashback radii and masses are computed by \sparta (see Section~\ref{sec:methods:sparta}).

We consider only the density profiles of isolated (or host) haloes because those of subhaloes typically contain a large contribution from the host. We generally define subhaloes as those that lie within $\rtombnd$ of another, larger halo (\citealt{behroozi_13_trees}, \citetalias{diemer_20_catalogs}), but we discuss the impact of other choices in Section~\ref{sec:methods:halo_sel}.

When considering the question of universality, we need to fairly compare halo masses across redshifts and cosmologies. For this purpose, we convert mass to peak height, $\nu_{\rm X}$, which captures the statistical significance of haloes in the linear overdensity field. It is formally defined as $\nu_{\rm X} = \deltac / \sigma(M_{\rm X})$, where $\deltac(z) = 1.686$ is the threshold overdensity in the top-hat collapse model \citep{gunn_72} and $\sigma(M_{\rm X})$ is the variance of the linear power spectrum measured in spheres of the Lagrangian radius of a halo, $R_{\rm L}$, the comoving radius that encloses the mass $M_{\rm X}$ at the mean density of the Universe, $M_{\rm L} = M_{\rm X} = (4 \pi/3) \rho_{\rm m}(z=0) R_{\rm L}^3$. We do not apply corrections for the finite volume of our simulations or for a slight evolution of $\deltac$ in \LCDM cosmologies \citep{mo_10_book} because these effects lead to tiny changes in $\nu$ that are relevant when computing mass functions but irrelevant for our purpose of binning haloes \citep{diemer_20_mfunc}. While we could compute peak height from any $M_{\rm X}$, we generally use $\nutomall$, which is denoted $\nutom$ or simply $\nu$ for brevity. We use the \colossus code \citep{diemer_18_colossus} to compute peak heights, approximating the power spectrum with the transfer function of \citet{eisenstein_98}. We refer the reader to \citet{diemer_20_mfunc} for a much more detailed description of these calculations. 

Finally, we define the logarithmic mass accretion rate per dynamical time following \citetalias{diemer_17_sparta}, using $\Gamma$ as a shorthand,
\begin{equation}
\label{eq:gammadyn}
\Gamma \equiv \gammadyn(t) \equiv \frac{\log[\mtom(t)] - \log[\mtom(t-\tdyn)]}{\log[a(t)] - \log[a(t - \tdyn)]} \,.
\end{equation}
Using mass definitions other than $\mtom$ makes little difference because the overall normalization is divided out, but the definition of the dynamical time does matter \citep{xhakaj_19_accrate}; we use the crossing time, $t_{\rm dyn} = 2 \rtom / \vtom$, where $\vtom = \sqrt{G\mtom / \rtom}$ \citep{diemer_17_sparta}. The relatively large radius of $\rtom$ reduces any baryonic effects on the accretion rate. 

\subsection{The SPARTA framework}
\label{sec:methods:sparta}

\sparta is a flexible, parallel C framework for the dynamical analysis of particle-based astrophysical simulations. Based on input from a halo finder (\rockstar and \consistenttrees in our case), the code tracks the trajectories of haloes and their constituent particles; the trajectory relative to the halo centre is called an orbit. The user can add plugin-style modules that analyse each orbit. The aggregated results can then be further analysed on a halo-by-halo basis. For example, \sparta determines the splashback radius by tracking all particles that enter each halo and measuring their first apocentre. From the locations and times of these events, it computes the splashback radius of a given halo by smoothing the distribution of particle splashbacks in time and taking its mean ($R_{\rm sp,mn}$) or higher percentiles (e.g., $R_{\rm sp,90\%}$ for the 90th percentile; \citetalias{diemer_17_sparta}). 

\vspace{-0.45cm}

\subsection{Orbit counting algorithm}
\label{sec:methods:oct}

\def\figsize{0.62}
\begin{figure*}
\centering
\includegraphics[trim =  5mm 6mm 3mm 2mm, clip, scale=\figsize]{\figdir/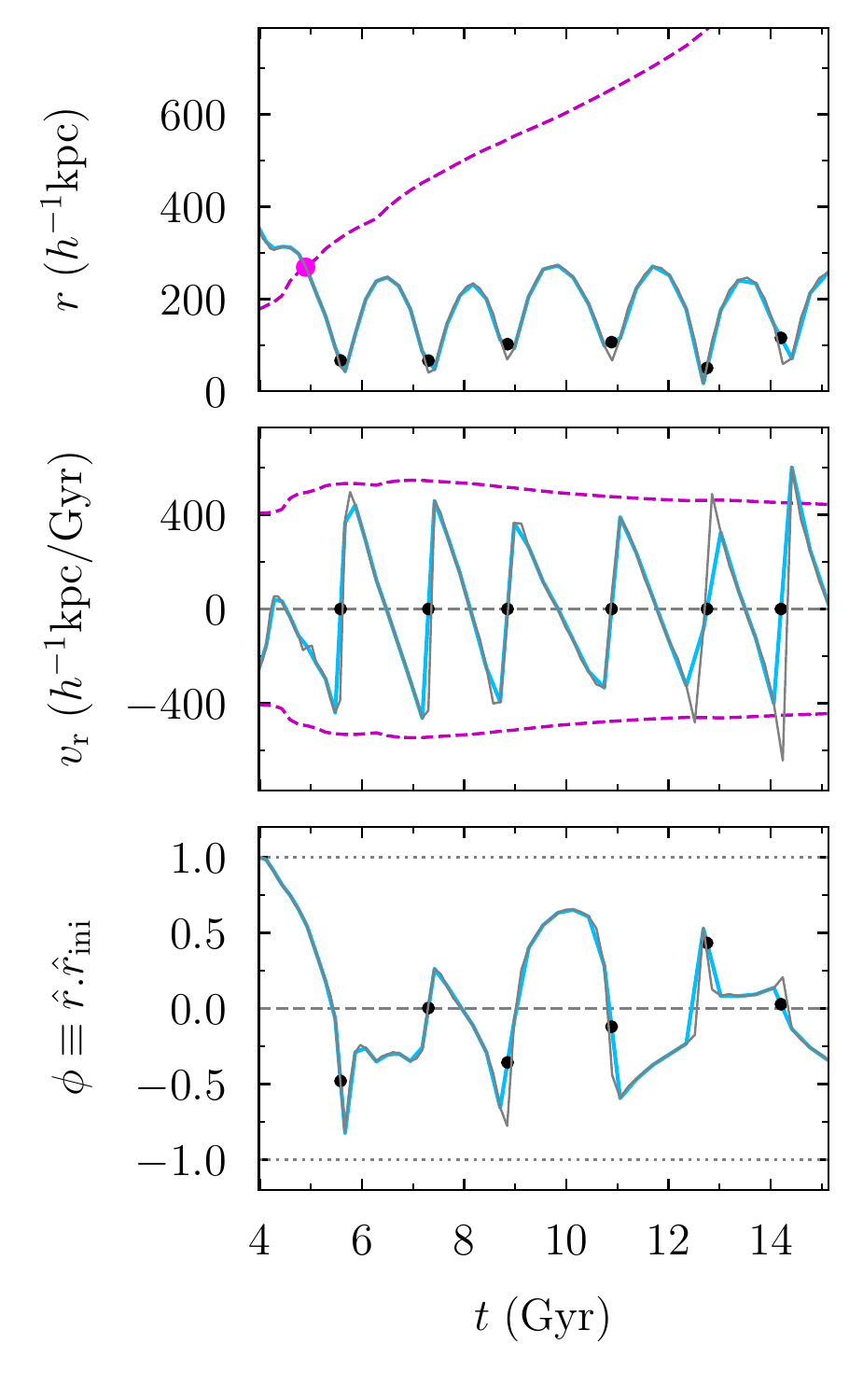}
\includegraphics[trim =  25mm 6mm 3mm 2mm, clip, scale=\figsize]{\figdir/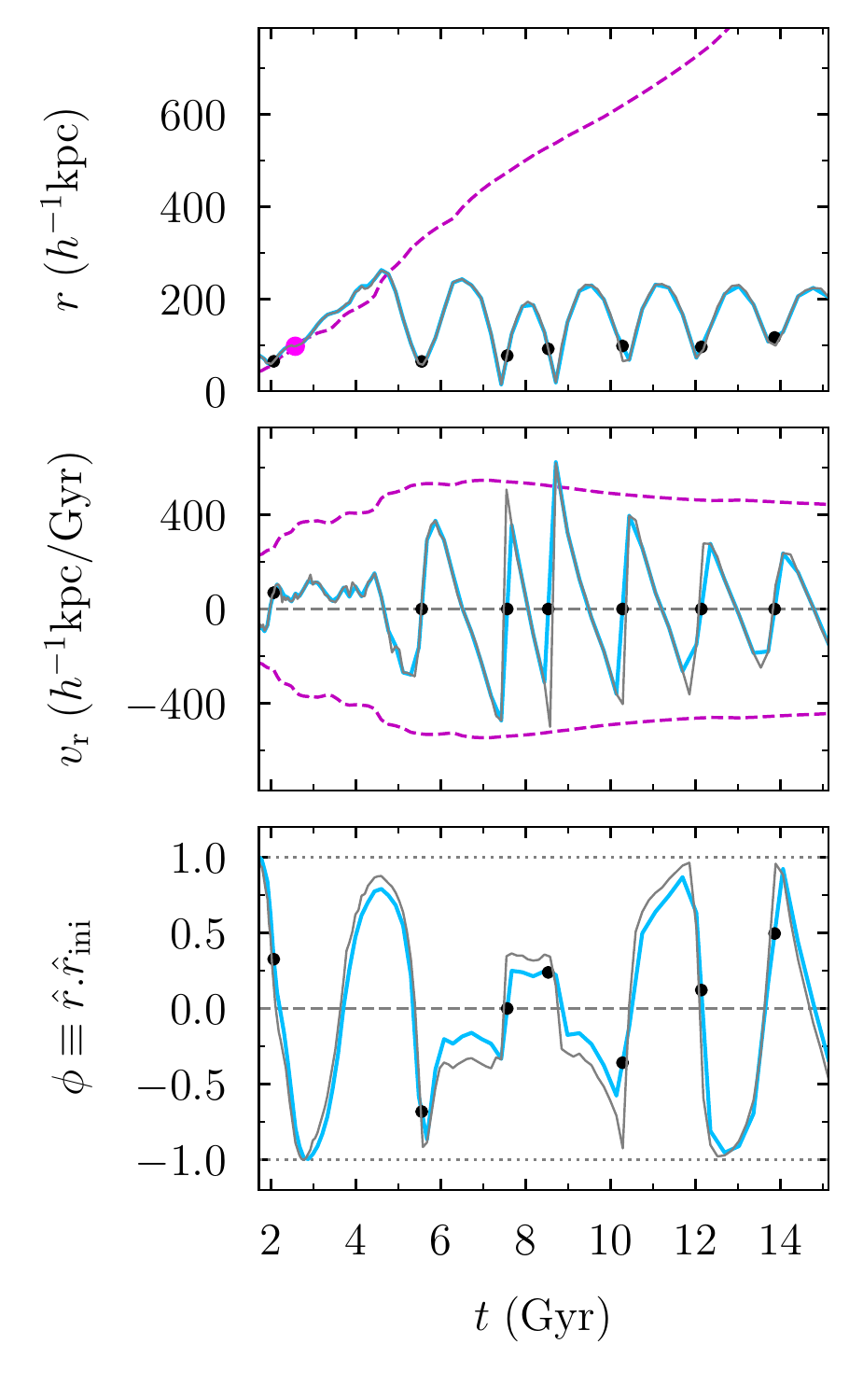}
\includegraphics[trim =  25mm 6mm 3mm 2mm, clip, scale=\figsize]{\figdir/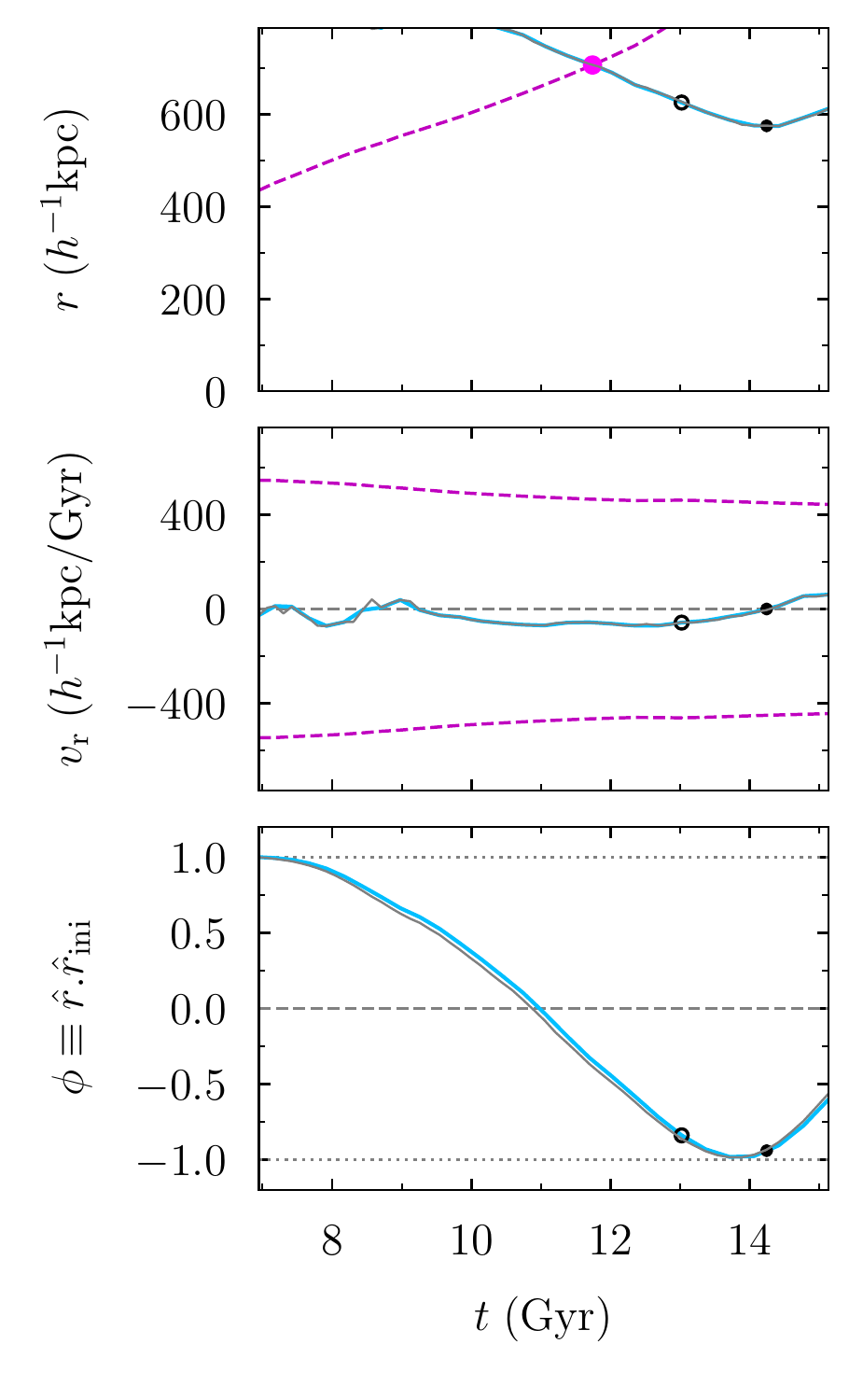}
\includegraphics[trim =  25mm 6mm 3mm 2mm, clip, scale=\figsize]{\figdir/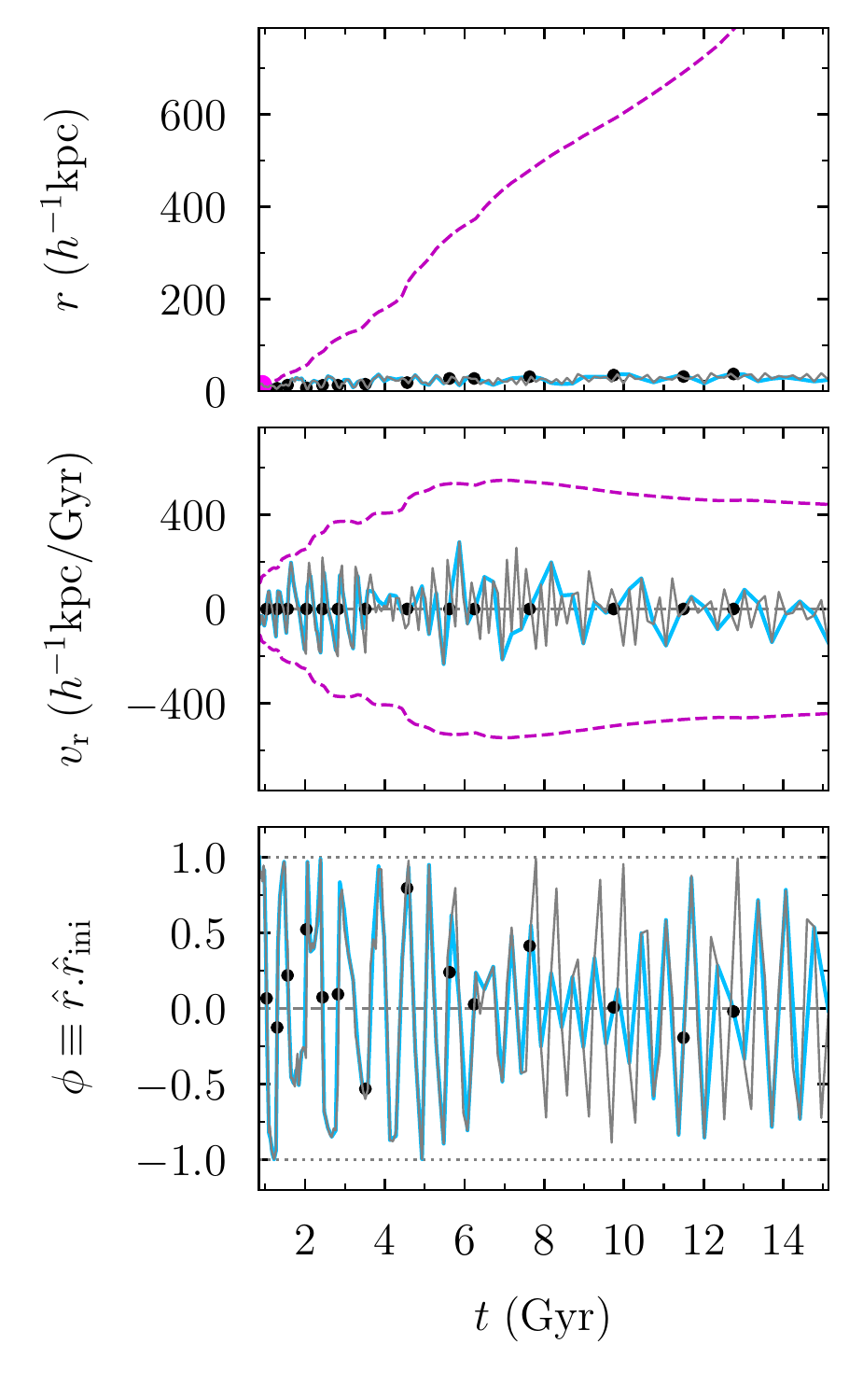}
\caption{Four typical particle orbits in an arbitrary halo in TestSim100 (blue) and TestSim200 (gray, with twice better time resolution). The rows show radius, radial velocity, and $\phi$, the dot product of the current and initial position vectors (which is unity at the beginning of the trajectory by construction). The dashed magenta lines show the evolution of $\rtom$ and of the circular velocity of the halo, $\vtom = \sqrt{G\mtom/\rtom}$. The magenta dots in the top panels indicate the time and radius where the particle first crosses $\rtom$. The black dots show the pericentres detected by our algorithm, where we have interpolated the pericentre time to coincide with $\vr = 0$. The first column shows a regular orbit with a roughly fixed apocentric distance. Our algorithm is not fooled by the short epoch with $\vr > 0$ before infall because $\phi \approx 1$ indicates that the particle has not changed its angular direction. Conversely, the algorithm does detect an early pericentre where $\phi \ll 1$ in the second orbit before the particle has even crossed $\rtom$. The third column shows a highly tangential orbit, which never comes close to the halo centre, exhibits little variation in the radial velocity, and smoothly transitions from alignment to anti-alignment. At about $13\ \gyr$, our algorithm sets a lower limit flag (open circles) because $\phi < -0.8$. In this particular case, a pericentre is also detected soon thereafter because the radial velocity does become slightly positive, but it is reassuring that the algorithm does not need to rely on this error-prone feature. The fourth column shows an orbit that is unresolved in time. While our algorithm correctly determines the first pericentre, many of the following pericentres cannot be identified in the noisy orbital data.}
\label{fig:orbits}
\end{figure*}

Our new orbit counting algorithm is implemented in the same framework but counts pericentres instead of apocentres. The goal is to determine whether each particle in a halo is orbiting (part of the one-halo term) or falling in for the first time. We choose a simple criterion: a particle is orbiting if it has undergone at least one pericentre, and infalling otherwise.\footnote{In reality, the infall term consists of both particles that are falling into the halo in a non-linear fashion and the correlated large-scale structure, which is often dubbed the `two-halo term'. However, there is no meaningful dynamical distinction between those terms because the turnaround radius of a halo grows with time, meaning that more and more matter begins to fall in. In a \LCDM cosmology, this process is eventually halted by the accelerating expansion of the Universe, but without running simulations far into the future, there is no way to decide which matter ends up in a halo \citep[][]{busha_05}.} This definition reduces our task to counting pericentres in the orbits of each particle. In principle, a pericentre should manifest itself as a switch from negative to positive radial velocity with respect to the halo centre and as a minimum in radius (see example orbits in \figmn{fig:orbits}). In practice, however, a reliable algorithm needs to wrestle with three sources of confusion: noise in the trajectories, particles that fall in as part of subhaloes, and highly tangential orbits. 

Before tackling these issues, we need to clarify our definition of a pericentre. Both theoretically and practically, it makes no sense to allow pericentres at arbitrarily far distances and long times before a particle enters a halo. For example, would we call a particle's motion an `orbit' if it resides at, say, $4 \rtom$? We could demand the particle to be gravitationally bound at the time of pericentre, but boundness is an ill-defined concept in cosmological simulations (see \citealt{behroozi_13_unbound} and the discussion in \citetalias{diemer_20_catalogs}). We do not introduce a maximum radius for pericentres because such limits lead to noticeable, unphysical features in the infalling density profile. Instead, we require that our algorithm must be able to discern between spurious and real pericentres even at large distances. In practice, however, we cannot measure pericentric radii greater than the radius where we begin to track the infall of particles, which we set to $2\ \rtom$ in our \sparta runs (\citetalias{diemer_17_sparta}, \citetalias{diemer_20_catalogs}; see Appendix~\ref{sec:tests:parameters}).

We generally avoid using radius information and instead focus on the radial velocity, $\vr$. All velocities are understood to be in physical units, meaning that they include the Hubble expansion flow. Besides a particle's velocity, we also consider its angular direction, which we quantify as the dot product of the current and initial unit position vectors, $\phi(t) \equiv \hat{\vect{r}}(t) \dotp \hat{\vect{r}}(t_{\rm ini})$. This quantity deviates from unity as the particle's angular position moves away from its direction when it was first discovered (at $2\ \rtom$) and reaches $-1$ at the opposite side of the halo (bottom panels of \figmn{fig:orbits}). We find that tracking particles only after they cross $\rtom$ (rather than $2\ \rtom$) can lead to a fairly different reference direction because it ignores the tangential motion that may have occurred previously (Appendix~\ref{sec:tests:parameters}).

As soon as a particle is being tracked, we begin looking for switches in the sign of $\vr$ from negative to positive. By itself, however, this criterion does not suffice. First, it picks up many spurious detections, e.g., due to halo finder noise in the centre positions of haloes. Second, in tangential orbits, the radial velocity may change very little during pericentric passages. Third, we expect particles to approach positive radial velocity {\it on average} as we approach the turnaround radius, which indicates that they are part of the Hubble flow rather than that they are orbiting the halo. To screen out unphysical pericentres, we demand that the first (but not any subsequent) pericentric passage should coincide with a switch in the angular position of the particle. For radial orbits, this switch occurs suddenly and in conjunction with the sign switch in $\vr$. For tangential orbits, the velocity switch may be gradual, but the angular direction slowly approaches anti-alignment with the original infall direction (\figmn{fig:orbits}). For the purposes of validating potential first pericentres, we demand that $\phi < 0.5$ at the snapshot before or after $\vr$ changes sign. This criterion is robust to small shifts in the radial and angular trajectories. For example, if $\vr$ becomes positive but $\phi$ is too close to unity, a pericentre will be triggered at a subsequent snapshot as long as $\phi$ keeps decreasing and $\vr$ stays positive (as is be the case for pericentres that are reasonably well resolved in time). 

By visual inspection, we find that this basic algorithm reliably identifies the vast majority of first pericentres. There are, however, a  few special cases. First, pericentres in nearly circular orbits may not be accompanied by a sign switch in $\vr$ due to noise in the trajectory (third column of \figmn{fig:orbits}). We identify this situation if no pericentre has been detected yet but $\phi$ falls below $-0.8$, that is, if the particle is almost exactly anti-aligned with its infall direction.\footnote{Technically, a pericentre should coincide with $\phi = 0$ rather than $\phi \approx -1$, the latter corresponding to an apocentre. However, we find that triggering the lower limit flag at $\phi = 0$ is too aggressive because this angular non-alignment frequently occurs slightly before a physical pericentre, as indicated by a switch in the radial velocity.} We cannot, however, detect the exact nature or time of such a pericentre and thus set a `lower limit' flag that indicates that the current pericentre count of zero may be incomplete. When constructing density profiles, we will add particles with this flag into the orbiting term regardless of their pericentre count. Second, we cannot find the first pericentre of particles that are already inside a halo when it is first detected by the halo finder. Again, we set the lower limit flag, meaning that the orbiting profile could erroneously include infalling particles for the first few snapshots during which a halo is alive. Conversely, particles that have orbited and are outside $\rtom$ when the halo is created are counted into the infalling population. Both problems are irrelevant in practice because the halo finder detects haloes with as few as $20$ particles while we analyse only haloes with at least $500$ particles (Section~\ref{sec:methods:halo_sel}), which will have existed for many dynamical times. A third challenging situation occurs when orbits are unresolved in time, i.e., if the orbital time is shorter than the snapshot spacing of the simulation. In this case, the conditions for a pericentre detection may never be fulfilled but the velocity tends to switch sign in a seemingly random fashion (right column in \figmn{fig:orbits}). We screen for such orbits by tracking the total number of times the velocity changes from negative to positive. We only count instances that are plausible pericentres, that is, if they occur within $\rtom$ and at least half a dynamical time (one radius-crossing time) after infall into the halo. If we find three such changes in velocity but have not yet detected a pericentre, we conclude that the orbit is unresolved and mark it as a lower limit. This case is so rare that makes no appreciable difference in the resulting profiles.

In summary, we record a first pericentre if we find a switch to positive velocity accompanied by $\phi < 0.5$ and a lower limit if $\phi < -0.8$, if a particle is found inside a newly discovered halo, or if we find three switches of the radial velocity. We have tested this algorithm extensively (Appendix~\ref{sec:tests}). We do not attempt to compute the exact time or radius of each pericentre because we are interested only in the first snapshot at which a particle is counted into the orbiting term. Our algorithm works with only three snapshots in time per orbit, which optimizes the memory footprint of the method. We have ensured that the outcome does not depend on the maximum radius to which particle trajectories are followed by \sparta. If a particle strays beyond this limit, its trajectory is interrupted but its orbit count is stored and reactivated in case the particle re-enters the halo.

The resulting orbit counts differ from halo to halo and depend on the simulation's box size and cosmology. We can get a general sense by considering the roughly $17$ million particles tracked in TestSim100 at $z \approx 0$ (that is, all particles that ever came within $2\ \rtom$ of a halo). Out of those, $15\%$ have orbited $N_{\rm orb} = 0$ times, out of which $12\%$ are marked as lower limits ($1.9\%$ of the total population). The majority of the lower limit particles are truly orbiting, meaning that the population of uncertain classifications is very small. The most populated bin is $N_{\rm orb} = 1$ with $38\%$ of the particles. As expected, the particle population in each orbit decreases in number thereafter, falling to less than $2\%$ at $N_{\rm orb} = 9$ (the highest number tracked). The lower limit fraction is below $1\%$ in all bins where $N_{\rm orb} > 1$. However, we emphasize that the results for $N_{\rm orb} > 1$ are not important for this paper and suffer from algorithmic uncertainties such as unresolved orbits (\figmn{fig:orbits}). It would be a challenging undertaking to obtain a numerically reliable, converged distribution of orbit counts at all $N_{\rm orb}$ \citep[see][for an attempt]{sugiura_20}.

\subsection{Dynamically split density profiles of individual haloes}
\label{sec:methods:prf}

Given the split into orbiting and infalling particles, we now construct the corresponding cumulative mass profiles (also within \sparta). We choose $80$ logarithmically spaced radial bins spanning the range $[0.01, 10]\ \rtomall$ (see Figs.~\ref{fig:viz1} and \ref{fig:viz2} for two examples). Ideally, we would extract profiles to even greater radii, but the runtime and memory consumption of \sparta increase steeply as the particle distribution stored on each process becomes the dominant factor. For example, the largest halo in the L0063-WMAP7 box has $\rtom = 1.8\ \mpch$ at $z = 0$, leading to an outer profile radius of $18\ \mpch$ or almost $0.3$ times the box size. 

One algorithmic subtlety is that we only find out about pericentres one snapshot after they have occurred (because we require two consecutive positive velocities, Section~\ref{sec:methods:oct}). Particles that have just undergone pericentre would be erroneously counted into the infalling term, an error that affects the profiles over a wide range of radii because particles can travel a significant fraction of $\rtom$ between snapshots \citepalias{diemer_17_sparta}. We correct this error by adding such particles to the orbiting term of the profile recorded in the previous snapshot. This correction, however, fails at the very last snapshot of the simulation because we cannot record pericentres that lie in the future. Thus, the last (usually $z = 0$) orbiting term is very slightly underestimated; we use the profile at the second-to-last snapshot ($z=0.03$ in our simulations) instead.

There are two numerical effects that render profiles unreliable at small radii, namely two-body relaxation and reduced centripetal forces due to force softening \citep[e.g.,][]{moore_98, klypin_01, power_03}. The recent calibration of \citet{ludlow_19} suggests that two-body relaxation should affect the profiles in the \wmap simulations by less than 5\% outside a minimum radius of $r_{\rm conv} = 0.086\ l(z)$, where $l(z) = L / N / (1 + z)$ is the inter-particle separation in physical length units (Table~\ref{table:sims}). The pre-factor changes to $0.091\ l(z)$ and $0.133\ l(z)$ for the \planck and self-similar simulations, respectively. Similarly, \citet{mansfield_21_resolution} suggest that force resolution should reduce the profiles by less than 5\% at radii larger than four force softening lengths ($4 \epsilon$, Table~\ref{table:sims}). In all plots and calculations, we exclude bins with a lower edge smaller than the more stringent of the two limiting radii. We present the detailed calculations that lead to our criteria, as well as quantitative tests, in Appendix~\ref{sec:tests}.

\subsection{Averaged density profiles}
\label{sec:methods:prf_av}

The density profiles of individual haloes are subject to significant random noise, as visualized in Figs.~\ref{fig:viz1} and \ref{fig:viz2}. The resulting scatter (quantified in Section~\ref{sec:results:scatter}) means that it is often difficult to derive meaningful conclusions from individual profiles. Thus, we will mostly consider the averaged (mean and median) density profiles of halo samples selected by certain properties. The radial bins in $r / \rtom$ typically contain different numbers of valid profiles because the radial resolution cut occur at different radii for different halos in a given sample, since they have different physical sizes and originate from different simulations. We compute the uncertainty on the mean and median by bootstrap subsampling the profile values in each bin $500$ times. The resulting uncertainty is typically very small due to the large number of profiles and due to their good numerical convergence (Appendix~\ref{sec:tests}).

A difficulty arises when computing the averages in regions where the number of particles per bin approaches zero. The mean and median statistics exhibit different issues in this case: the mean gradually becomes noisier, whereas the median density may suddenly fall to zero. To avoid such spurious results, we compute the expected number of particles in a radial bin based on the mean $\rho / \rhom$ profile of the overall halo sample, from the bin volume in each individual halo, and from the particle mass of the given simulation. We drop the radial bin from the mean profile of the sample if it contains fewer than $1000$ particles total. Technically, this criterion introduces a bias because we used the mean profile itself to estimate the number of particles, but the statistical fluctuation scales as $1/\sqrt{N}$ and is thus negligible for a $1000$ particle limit. For the median profiles, we exclude individual halo profiles from a radial bin if their expected particle count falls below $10$. This method is unbiased because it relies on the sample's mean profile rather than on the individual halo's profile. We have verified that our limits remove spurious medians and noisy bins without biasing the results. 

In principle, we might want to rescale profiles by a radius other than $\rtomall$, which would necessitate an interpolation to a new radial grid. In this paper, however, we restrict ourselves to profiles scaled by $\rtomall$ because the transition region is most universal in this radius, at least compared to other spherical overdensity definitions \citepalias{diemer_14}.

\subsection{Halo selection and resolution limits}
\label{sec:methods:halo_sel}

Besides omitting unreliable radial bins, we also exclude entire halos that are insufficiently resolved or otherwise biased. First, we consider only haloes with $N_{\rm 200m,all} \geq 500$. This limit is slightly less strict than that of \citetalias{diemer_14}, but we find good convergence (Appendix~\ref{sec:tests}) in agreement with \citet{mansfield_21_resolution}, who showed that profile-related quantities such as $\vmax$ are converged at $500$ particles in the \erebos simulations.

Second, we include only host (isolated) haloes because the density profiles of subhaloes are ill-defined due to the (often dominant) contribution from the host. However, the distinction is ambiguous given different definitions of the halo boundary \citep{diemer_21_subs}. We have experimented with the host-subhalo relations from the halo catalogues of \citetalias{diemer_20_catalogs}. Using a small radius such as $\rfoc$ means including haloes that reside well within $\rtom$ of larger haloes, and thus higher mean and median densities at large radii. Conversely, using a large radius such as $R_{\rm sp,90\%}$ lowers the average outer profiles significantly. Physically speaking, we are not interested in the contributions from other haloes as they are a manifestation of the halo--halo correlation function as much as of the density structure around isolated haloes. These contributions mostly affect the infall term of low-mass haloes because high-mass haloes are less likely to reside near another halo of comparable or larger size. We can thus gauge the effect of neighbouring haloes by comparing the mean or median profiles of samples with different peak height, and we find that excluding subhaloes out to larger radii brings the profiles into better agreement. On the other hand, using a radius as large as $R_{\rm sp,90\%}$ defines a significant fraction of all haloes as subhaloes \citep{diemer_21_subs}.

As a compromise, we use $\rtombnd$ as our fiducial definition of subhaloes but introduce another criterion to screen for haloes with a large neighbour contribution: the fraction of particles that are not gravitationally bound. In particular, we exclude haloes for which  $M_{\rm 200m,all} / M_{\rm 200m,bnd} > 1.5$, although we emphasize that there is no `correct' value for this cut (Appendix~\ref{sec:tests}). Our choice represents a compromise that ensures relatively minor neighbouring-halo effects and excludes a small fraction of haloes: 2\% of all haloes in the \wmap sample at $z = 0$, up to 5\% at higher redshifts, and similar fractions in the self-similar simulations. This cut mostly affects haloes with high accretion rates. 

In summary, we include only haloes with $\ntom \geq 500$ particles and $M_{\rm 200m,all} / M_{\rm 200m,bnd} \leq 1.5$. Out of their profiles, we include only bins with radii $r > {\rm max}(4 \epsilon, r_{\rm conv})$ from the halo centre. In all figures, we omit radial bins that contain fewer than $80$ individual halo profiles, a compromise between showing profiles to small radii and hiding noisy bins that distract the eye.

\subsection{Halo samples across simulations and redshifts}
\label{sec:methods:combined}

We now combine the profiles of haloes from different simulations into one sample per cosmology and redshift. These samples can then be further split by mass, mass accretion rate, and so on. For the \LCDM simulations, we have computed profiles at $z = [8$, 7, 6, 5, 4, 3, 2, 1.5, 1, 0.8, 0.6, 0.5, 0.4, 0.3, 0.2, 0.1, 0.03]. Some of the highest redshifts have been omitted for the largest box sizes, where no haloes exist yet at early times (Table~\ref{table:sims}). In Appendix~\ref{sec:tests:res_tests}, we check that these samples are converged across simulations with different mass and force resolutions. The combined \wmap halo sample contains about \num{510000} haloes at $z = 0$, decreasing to \num{38000} at $z = 6$. The \planck sample contains \num{320000} haloes at $z = 0$.

For the self-similar simulations, we combine different redshifts into one sample per simulation because cosmic time is not physically meaningful. We compute profiles at redshifts that correspond to even spacings in dynamical time \citep{diemer_20_mfunc}, namely, for $n = -1$ at $z = [$15, 11.5, 8.8, 6.8, 5.1, 3.8, 3.3, 2.8, 2.3, 2.085], for $n = -1.5$ at $z = [$12.4, 9.6, 7.3, 5.6, 4.2, 3.1, 2.2, 1.5, 1.01], for $n = -2$ at $z = [$11.7, 9.0, 6.9, 5.2, 3.9, 2.9, 2.1, 1.4, 0.9, 0.531], and for $n = -2.5$ at $z = [$4.3, 3.2, 2.3, 1.6, 1.0, 0.6, 0.27, 0.03]. The profiles exhibit the expected self-similarity to an accuracy of about 10\% (Appendix~\ref{sec:tests:res_tests}). The combined halo samples vary in size from $1.9$ million for $n = -1$ to \num{180000} for $n = -2.5$.

As an example, \figmn{fig:prof:example} shows the averaged halo profiles of the \wmap sample at $z = 0$. The remaining profile figures follow the same scheme, where the three larger panels show the profiles of all, orbiting, and infalling matter ($\rhotot$, $\rhoorb$, and $\rhoinf$, from top to bottom). Below those panels, we show the logarithmic slope, $d \ln \rho / d \ln r$, which is computed using a 4th-order \citet{savitzky_64} filter with a window size of $15$ bins. This filter smooths over noisy bins without changing the physical features of the profiles \citepalias{diemer_14}. The Savitzky-Golay filter does not work for the $7$ left and rightmost bins (half the window size), where we estimate the derivative with a Gaussian smoothing as described in Appendix B of \citet[][using a window width of $\Delta_{\rmR} = 0.2$]{churazov_10}.

%%%%%%%%%%%%%%%%%%%%%%%%%%%%%%%%%%%%%%%%%%%%%%%%%%%
% RESULTS
%%%%%%%%%%%%%%%%%%%%%%%%%%%%%%%%%%%%%%%%%%%%%%%%%%%

\section{Results}
\label{sec:results}

In this section, we discuss the shapes of the orbiting, infalling, and total profiles. We focus on averaged profiles because those of individual haloes are subject to significant random noise. We consider haloes across vast ranges of mass, redshift, and cosmology, and combine them into samples binned by mass, redshift, and accretion rate. Given the large parameter space, we summarize the most important trends with a few representative examples and refer the curious reader to a more comprehensive selection of online figures (Section~\ref{sec:intro}). In particular, we discuss the profiles of halo samples selected by only mass and redshift in Section~\ref{sec:results:nu} and \figmn{fig:av:nu}. We additionally bin haloes by accretion rate in Section~\ref{sec:results:nugamma} and \figmn{fig:av:nugamma:md}. We study the impact of cosmology in Section~\ref{sec:results:cosmo} and \figmn{fig:av:cosmo}. And finally, we consider halo-to-halo scatter in Section~\ref{sec:results:scatter} and \figmn{fig:scatter}.

\subsection{Haloes selected by peak height}
\label{sec:results:nu}

\begin{figure}
\centering
\includegraphics[trim =  5mm 8mm 0mm 2mm, clip, width=0.43\textwidth]{\figdir/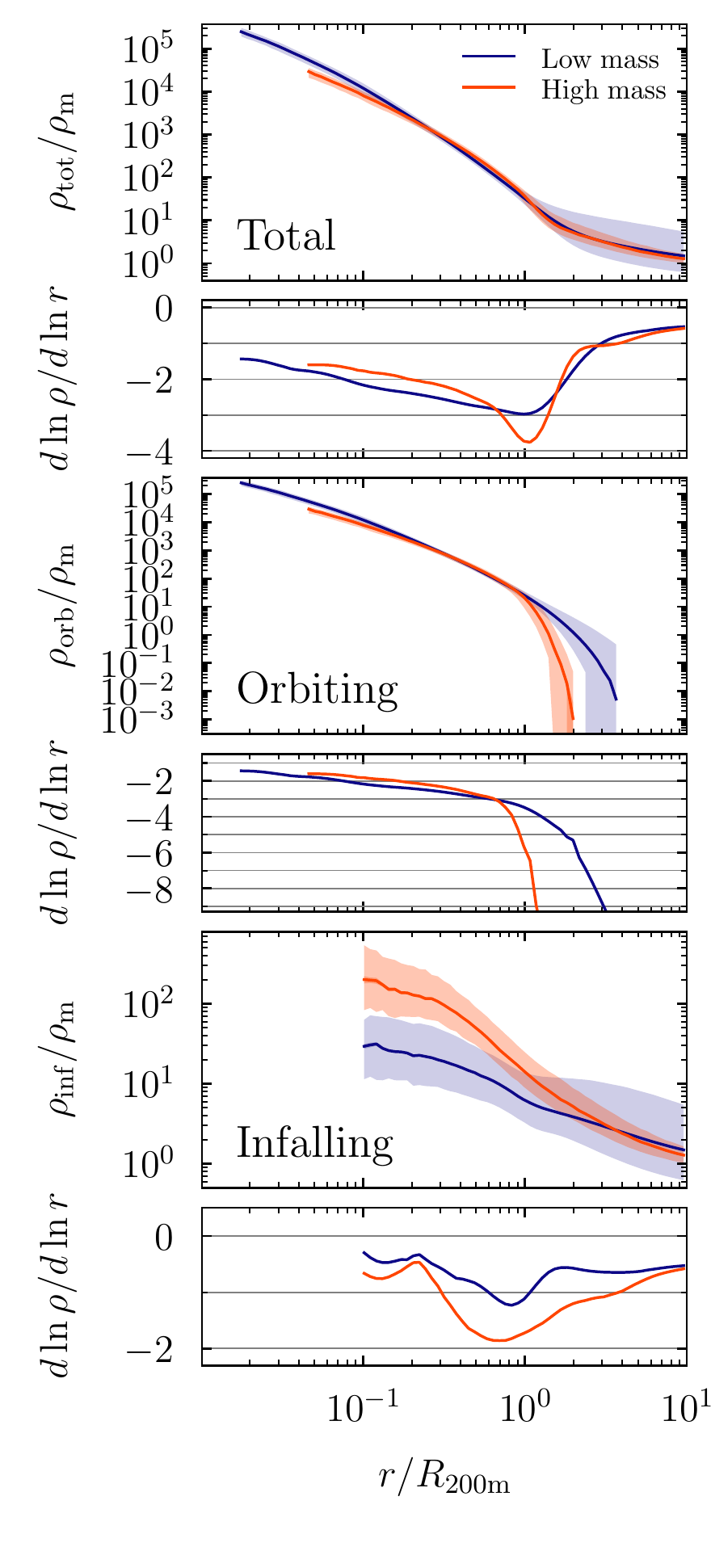}
\caption{Example of dynamically split profiles for two \LCDM halo samples. Each panel shows the median profiles (lines) and 68\% scatter (shaded areas) for low-mass haloes ($0.5 < \nu < 1$, blue) and high-mass haloes ($\nu > 3$, orange) in the \wmap cosmology at $z = 0$. The larger three panels show the total, orbiting, and infalling profiles; the smaller bottom panels show the respective logarithmic slopes, $\dnorminl{\ln \rho}{\ln r}$. The remaining figures in this paper are arranged similarly. The halo-to-halo scatter in the slopes is not shown because it is difficult to compute the slope of individual profiles in a robust manner. The example shown highlights a number of the most important trends. The scatter is larger around low-mass haloes because they are more influenced by nearby neighbours (Appendix~\ref{sec:tests:neighbours}). The median profiles of low-mass haloes steepen gradually, while those of high-mass haloes exhibit a clear edge at the distribution of orbiting matter. The position and sharpness of this feature depend on mass via the different mass accretion rates \citepalias{diemer_14}. While the total profiles reach slopes of about $-4$, the slope of the orbiting profile becomes arbitrarily steep at the truncation radius. The infalling profiles also vary systematically with halo mass and exhibit significant halo-to-halo scatter, especially for low-mass haloes.}
\label{fig:prof:example}
\end{figure}

\begin{figure*}
\centering
\includegraphics[trim =  3mm 9mm 0mm 2mm, clip, width=\textwidth]{\figdir/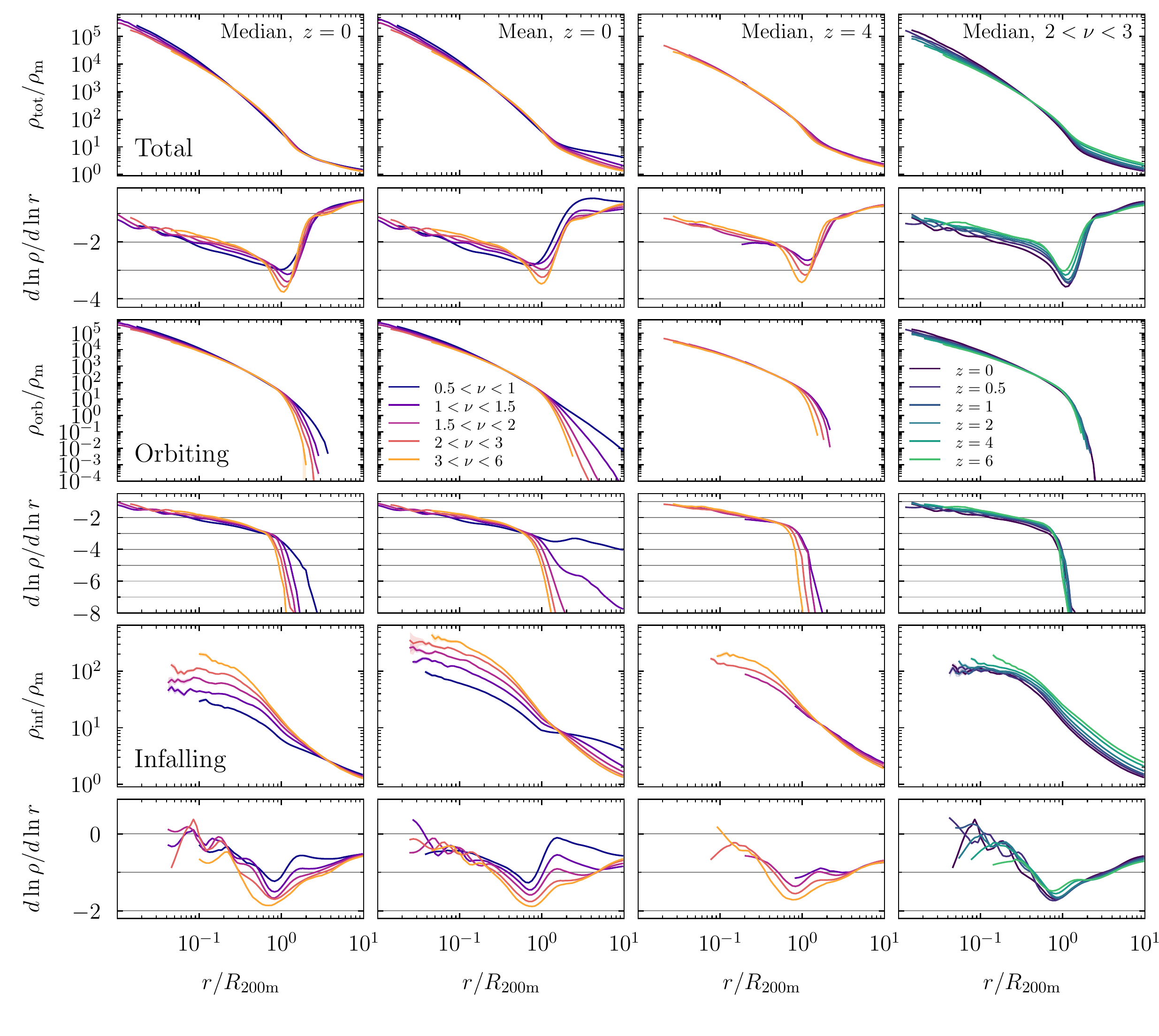}
\caption{Average density profiles of haloes in the \wmap cosmology, selected by peak height and redshift. As in \figmn{fig:prof:example} and in all following figures, the large panels show the total, orbiting, and infalling profiles (from top to bottom). In the left column, we compare median profiles at $z = 0$ across masses (from $\nu = 0.5$ or $\mtom > 2 \times 10^{10}$ in blue to $\nu > 3$ or $\mtom > 1.1 \times 10^{15}\ \msun$ in yellow). The truncation of the orbiting term (the splashback radius) occurs at smaller radii in higher-mass samples because they also have higher mass accretion rates on average. The infalling profiles also evolve with mass, mostly for a similar reason (Section~\ref{sec:results:nugamma}). The second column shows the mean profile of the same samples. In low-mass halos, the mean orbiting term reaches larger radii because of a few haloes where matter has been removed to large radii by interactions with other haloes. The third column shows the median profiles at $z = 4$, which appear fairly similar to their $z = 0$ counterparts. We confirm this impression in the fourth column, where we track the evolution of the median profiles for $2 < \nu < 3$ haloes across redshift (the trends are representative for other mass bins as well). The redshift evolution of the orbiting term is modest, but the normalization of the infalling term increases, leading to an overall shallower slope in the total profiles.}
\label{fig:av:nu}
\end{figure*}

We begin by binning haloes in \LCDM cosmologies in redshift and mass, which we express as peak height to obtain a redshift-independent measure of the relative size of haloes (Section~\ref{sec:methods:defs}). As demonstrated in \figmn{fig:prof:example}, we visualize our results as the total, orbiting, and infalling profiles from top to bottom, with smaller panels showing their logarithmic slope. We follow this pattern in \figmn{fig:av:nu}, which summarizes our results for mass-selected profiles in the \wmap cosmology. All conclusions of this section equally apply to the \planck cosmology, whose profiles are visually indistinguishable. We focus on the \wmap simulations because they cover a larger range of box sizes and halo masses.

In the left column of \figmn{fig:av:nu}, we compare the median profiles of halo samples at $z = 0$ from small galaxy haloes ($\nu = 0.5$ or $\mtom \approx 2 \times 10^{10}\ \msun$) to massive clusters ($\nu > 3$ or $\mtom > 1.1 \times 10^{15}\ \msun$). At higher redshift, a fixed peak height corresponds to lower mass. At first sight, the profiles look similar across the large mass range, but this impression is partly caused by the large dynamical range. The slope panels demonstrate that the shape of the profiles does depend on mass. While the total profiles of low-mass haloes smoothly steepen out to about $\rtom$, those of massive haloes are slightly shallower at small radii but then steepen significantly near $\rtom$ (as shown by \citetalias{diemer_14} and subsequent works). At larger radii, the profiles approach the shallower slope of the infalling profile.

With our dynamical split, we can now identify the pattern that leads to the profile shape in the transition region. The orbiting term experiences a sharp truncation at a radius that varies with mass (or rather, average mass accretion rate as we show in Section~\ref{sec:results:nugamma}). The logarithmic slope approaches arbitrarily steep values at this radius. Given the rapidly decreasing density of orbiting matter, the infalling term quickly comes to dominate, leading to a thin transition zone where the slope sharply drops and recovers. The infalling term itself depends on peak height, with higher mass haloes having steeper infalling profiles (again largely an effect of the average accretion rate). Moreover, the infalling profiles seem to approach different asymptotic density values at small radii, though the profiles tend to become unreliable (and are thus not plotted) before they clearly converge to a fixed density.

The radius where the slope is steepest is often taken to be a proxy for the splashback radius, but the profiles in \figmn{fig:av:nu} highlight why this is an imperfect definition: as both the orbiting and infalling terms depend on mass, the radius where the slope is steepest is only approximately equal to the radius where the orbiting term truncates. The difference is borne out when comparing the radius of steepest slope to dynamically calculated splashback radii (\citetalias{diemer_20_catalogs}, see also \citealt{oneil_22}). The truncation radius could be taken as a new proxy for $\rsp$, but we note that it is defined by all orbiting particles, whereas the splashback radius is defined by the apocentres of only the most recently accreted particles \citep{bertschinger_85, more_15}. We will explore the detailed connection between the truncation and splashback radii in \paperthree.

Our conclusions have been based on median profiles, but observations generally constrain mean profiles second column of \figmn{fig:av:nu}). We expect differences between mean and median for asymmetric distributions, where outlier profiles can strongly influence the mean. Indeed, we notice significant differences in the mean orbiting and infalling profiles of low-mass haloes. The orbiting term continues past the truncation radius that is apparent in the median profiles because a few disrupted haloes have significant `orbiting' material at large radii, which disproportionately contributes to the mean value at radii where the orbiting overdensity is small (say, below $\rhom$). Similarly, the mean infalling profile of the lowest mass sample exhibits a sharp plateau feature near $\rtom$. We suspect that this feature is related to the definition of the halo boundary because excluding haloes with large fractions of unbound material reduces it substantially (Appendix~\ref{sec:tests:neighbours}). Moreover, the feature occurs only in low-mass haloes that are more strongly influenced by their environment than their high-mass counterparts. Luckily, most observations of density profiles currently target group and cluster scale haloes with $\nu \gsim 2$, where the mean and median profiles agree much better.

\begin{figure*}
\centering
\includegraphics[trim =  3mm 9mm 0mm 2mm, clip, width=\textwidth]{\figdir/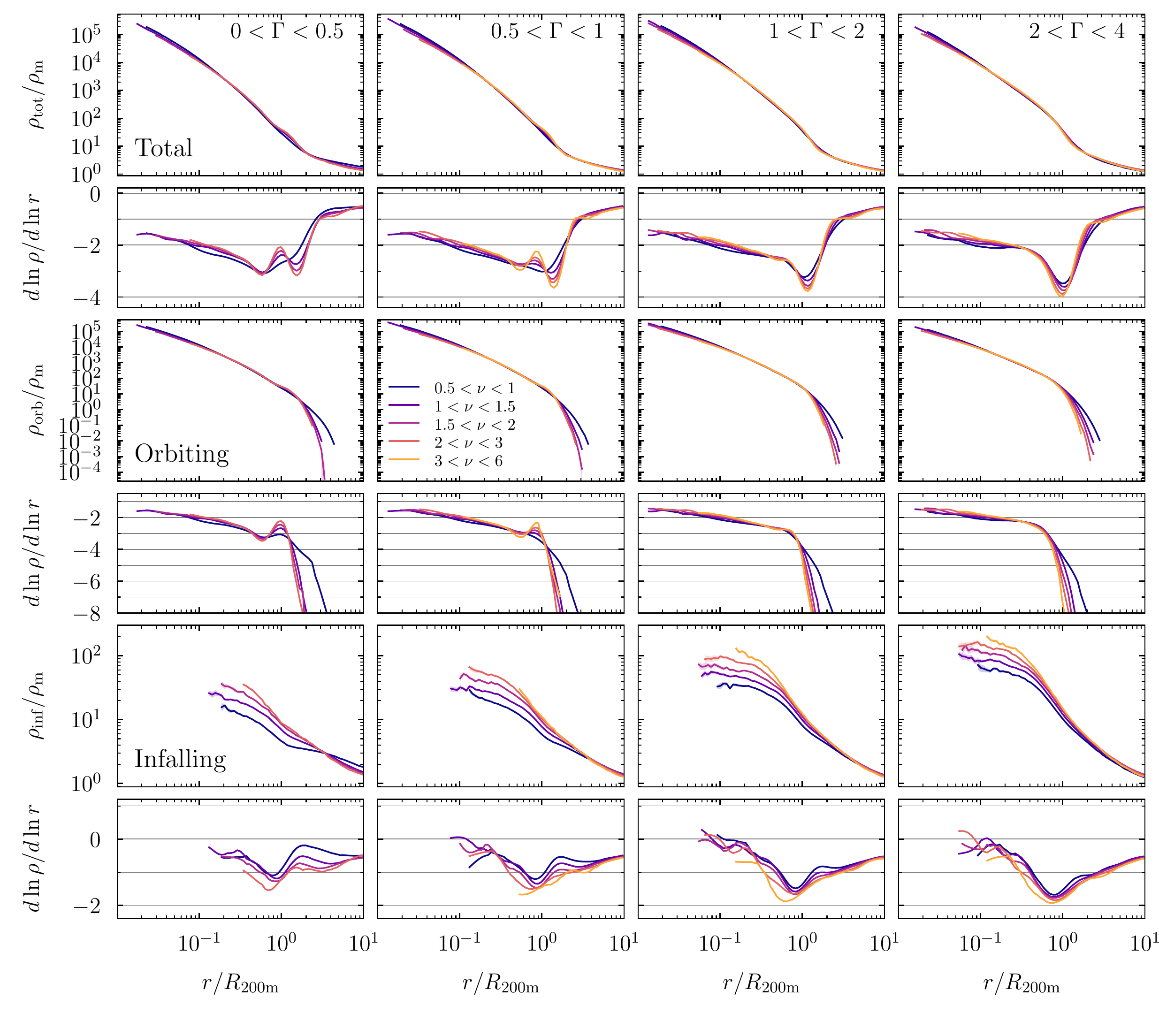}
\caption{Median profiles of samples split by both peak height (colours) and mass accretion rate (columns) at $z = 0$ in the \wmap cosmology. All trends are similar for other redshifts and cosmologies. Comparing to the profiles selected by only $\nu$ in the left column of \figmn{fig:av:nu}, a number of salient differences are readily apparent. The accretion rate is a stronger determinant of profile shape than peak height, given that the profiles in each column vary only mildly with $\nu$ but significantly across columns. The splashback feature becomes more pronounced with mass accretion rate and moves towards smaller radii. The inner profiles of slowly accreting halos exhibit smoothly decreasing slopes, whereas fast-accreting profiles resemble power laws. The wiggles in slope near the truncation radius in slowly accreting haloes (left two columns) are due to the second orbit of particles \citep{adhikari_14}. The shape of the infalling term is also strongly influenced by $\Gamma$, with a more variable slope in fast-accreting haloes.}
\label{fig:av:nugamma:md}
\end{figure*}

In the third column of \figmn{fig:av:nu} we investigate the redshift dependence of the profiles by comparing to haloes of the same peak heights but at $z = 4$. The lowest $\nu$ bins are not accessible at this redshift because they correspond to haloes below the resolution limit of our simulations. In general, the profiles appear similar, including the trends with peak height. For a more direct comparison between redshifts, we plot the $2 < \nu < 3$ profiles for a large range of redshifts in the right column (the evolution of the other $\nu$ bins proceeds similarly). Overall, the profiles become slightly shallower with redshift at all but the largest radii. However, the orbiting term remains remarkably unchanged with redshift \citep[see also][]{lebrun_18}, meaning that the evolution of the total profiles is driven by an increase in the normalization of the infalling profiles. Since all profiles are expressed in units of $\rho / \rhom$, this increase in normalization goes beyond the cosmological increase of $\rhom(z)$ with redshift and turns out to be related to the power spectrum (Section~\ref{sec:results:cosmo}).

\subsection{Haloes selected by mass accretion rate}
\label{sec:results:nugamma}

Most of the trends with halo mass are actually consequences of the dynamical state of haloes, namely, of their mass accretion rate $\Gamma$ (\eqmnb{eq:gammadyn}). Thus, profiles are most universal across mass, redshift, and cosmology when they are binned by $\Gamma$. \figmn{fig:av:nugamma:md} shows the median profiles for the same $\nu$ bins as in \figmn{fig:av:nu} but additionally split into four bins in $\Gamma$. The halos in the lowest bin barely accrete (or even lose) matter because the pseudo-evolution due to the evolving overdensity threshold of $200 \rhom(z)$ corresponds to roughly $\Gamma \approx 0.5$ \citep{diemer_13_pe}. The profiles for haloes with $\Gamma > 4$ are very similar to those in the $2 < \Gamma < 4$ bin, in agreement with \citet{diemer_17_rsp}, who found that $\rsp / \rtom$ approaches a constant value at high $\Gamma$. We only show results for \wmap and $z = 0$, but the trends are the same for the \planck cosmology and for higher redshifts.

\figmn{fig:av:nugamma:md} largely confirms the central conclusions of \citetalias{diemer_14}, which is not surprising given that they were based on the same simulations. Most notably, the truncation radii of the orbiting term (or splashback radii) are smaller for higher $\Gamma$. This trend arises because a particle with a given kinetic energy at infall will reach apocentre at a smaller radius if mass is added during its orbit. This effect is independent of halo mass \citep{more_15}, as evidenced by the alignment of the truncation radii for different $\nu$. One exception is the lowest $\nu$ bin (blue), where neighbouring haloes strongly influence the profiles. Conversely, the truncation appears sharpest in the most massive halos because they dominate their environment. All of these trends makes sense in the context of spherically symmetric infall models, where the boundary of the orbiting term is delineated by a single, well-defined shell of particles. We review such models in Section~\ref{sec:discussion:truncation}.

The inner profile ($r < \rvir$) also changes significantly with $\Gamma$. In slowly accreting halos, the profile slope decreases smoothly with radius, whereas the profiles of fast-accreting halos approach a power law with a slope of about $-2$ across a wide range of radii. Such a profile is a natural prediction of fast collapse with purely radial orbits (Section~\ref{sec:discussion:orb_slope}). In slowly accreting halos ($\Gamma < 1$, left two columns), we observe wiggles in the slope at $r < \rtom$, which reflect a second caustic caused by the second orbit of recently accreted particles \citep{adhikari_14}. In high-$\Gamma$ halos, this signature is washed out by particles on their first orbit.

Based on our new algorithm, we can for the first time discern the effects of mass and accretion rate on the infalling profiles. At low $\Gamma$, they are close to power laws (fixed slope) with a shallower slopes in smaller halos. When there is little physical accretion, environmental effects matter: the profiles of small halos contain significant contributions from close neighbours (see also \figmn{fig:conv2} and Appendix~\ref{sec:tests:neighbours}). At high $\Gamma$, the profiles become virtually independent of $\nu$ because they are driven by freshly accreted matter. Those profiles exhibit a characteristic shape where they steepen as matter approaches the truncation radius, although this trend mostly disappears when considering overdensity rather than density (\papertwo). The profiles flatten again as particles reaches the centre of the halo and appear to asymptote to a limiting central density at small radii, which is higher for higher $\nu$ haloes. To our knowledge, these features of the infalling profiles have not been predicted by theoretical models, which we further discuss in Section~\ref{sec:discussion:shape_inf}.

\subsection{The effect of cosmology and the power spectrum}
\label{sec:results:cosmo}

\begin{figure*}
\centering
\includegraphics[trim =  3mm 9mm 0mm 2mm, clip, width=\textwidth]{\figdir/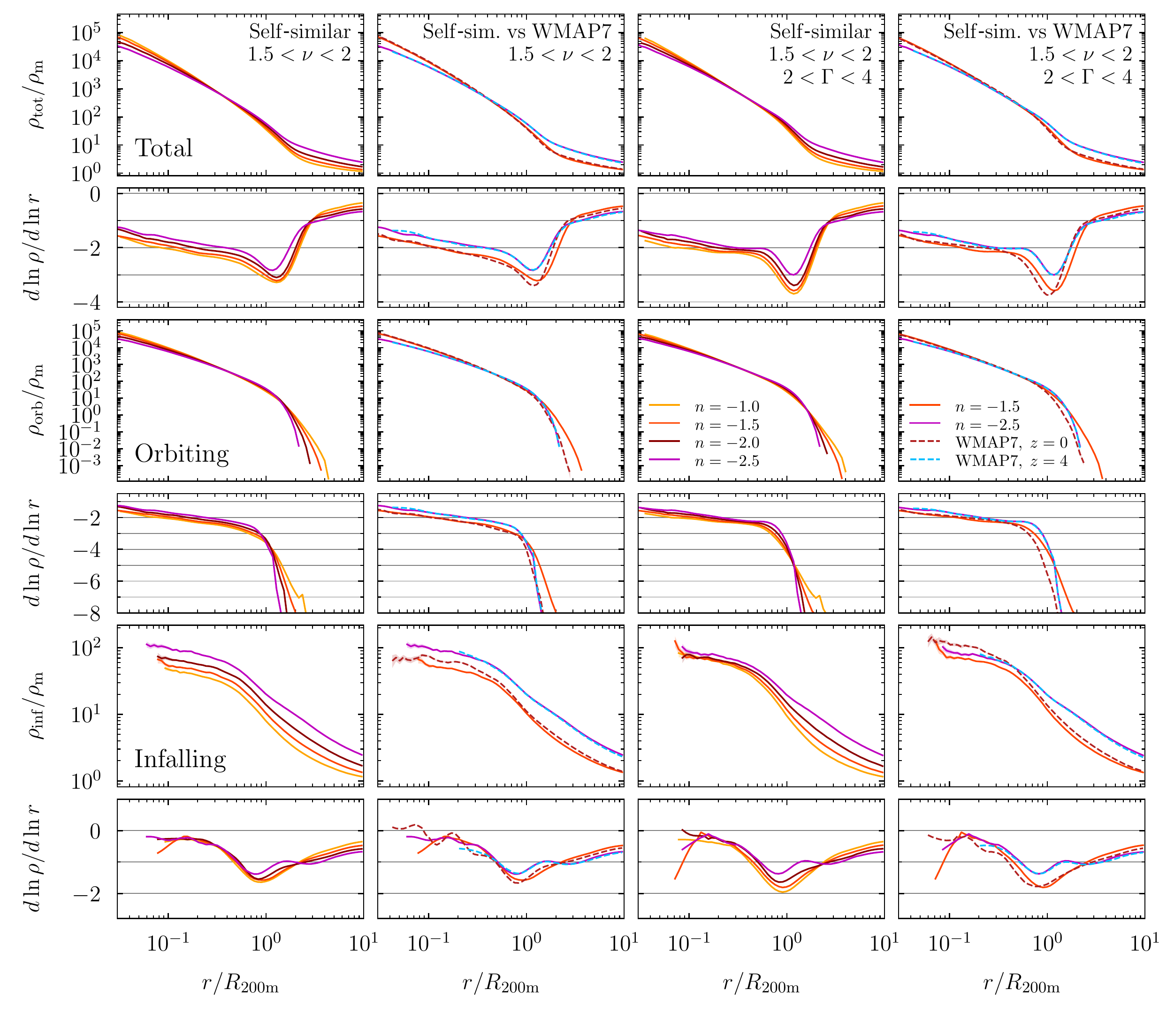}
\caption{Summary of the effect of cosmology on profiles. In each column, we compare the median profiles for haloes with $1.5 < \nu < 2$ for different cosmologies. In the right two columns, we additionally select by mass accretion rate and only include haloes with $2 < \Gamma < 4$. In the first and third columns, we compare profiles from the four self-similar simulations with different power spectrum slopes $n$. In the second and fourth columns, we repeat the lines for $n = -1.5$ (orange) and $n = -2.5$ (magenta) but contrast them with the \wmap \LCDM cosmology at $z = 0$ (red dashed) and $z = 4$ (blue dashed). These comparisons are exemplary of a number of general trends. The slope of the self-similar profiles is systematically shallower for steeper slopes of the power spectrum. This trend holds even when selecting by $\Gamma$ and affects both the orbiting and infalling terms. The \LCDM profiles at $z = 0$ are strikingly similar to $n = -1.5$ and those at $z = 4$ to $n = -2.5$, which demonstrates that much of the redshift evolution in \LCDM is due to the different part of the power spectrum probed by different halo masses.}
\label{fig:av:cosmo}
\end{figure*}

Having considered the impact of mass and accretion rate, we now turn to the effects of cosmology. Our unit system aims to be as inherently cosmology-independent as possible: peak height takes into account the varying linear growth factors in different cosmologies, we plot $\rho / \rhom(z)$ to take advantage of a natural scaling with the mean matter density, and $\Gamma$ is a logarithmic accretion rate that scales out the zeroth-order cosmology dependence of halo mass. By `effect of cosmology', we thus mean differences due to cosmological parameters at fixed peak height and redshift.

Comparing our \wmap results from Sections~\ref{sec:results:nu} and \ref{sec:results:nugamma} to the \planck cosmology, we find the orbiting terms to be statistically indistinguishable. The normalization of the infalling term appears to be a few percent lower in the \planck cosmology, regardless of halo selection. We can exclude a linear scaling with $\Omega_\rmm$ because the change in that parameter is $0.32 / 0.27 - 1 \approx 18\%$, much larger than the differences we find. However, the change in $\sigma_8$ and other parameters might partially offset the effects of the matter density. We also note that the matter--matter correlation function, $\xi_{\rm mm}(r)$, is higher (rather than lower) in the \planck cosmology at the relevant radii. Without a larger range of simulated cosmologies, we cannot present any firm conclusions regarding the dependencies on $\Omega_\rmm$ and $\sigma_8$.

We can, however, test the origin of differences due to cosmology more generally by considering Einstein-de Sitter universes. As structure formation is perfectly self-similar in these simulations, the profiles at fixed peak height agree across cosmic time (Appendix~\ref{sec:tests:res_tests}), leaving only one free parameter: the slope of the power-law initial power spectrum, $n$. While self-similar universes are not realistic, the dependence of halo properties on $n$ is a proxy for \LCDM cosmologies, where $n$ varies with spatial scale (and thus with mass and redshift). This correspondence has been demonstrated for concentration \citep{diemer_15, ludlow_16}, the abundance of subhaloes \citep{diemer_21_subs}, the slope evolution of Einasto profiles \citep{ludlow_17, udrescu_19}, and (phase-space) density profiles in general \citep{brown_20}.

The left column of \figmn{fig:av:cosmo} shows the median profiles in four self-similar universes with slopes between $n = -1$ and $-2.5$. The general shape of the profiles is very similar to \LCDM. The main effect of a shallower $n$ is to cause slightly steeper orbiting profiles out to the truncation radius and vice versa, a trend compatible with the literature \citep{crone_94, reed_05_profiles, knollmann_08, ludlow_17}. The slope of the infalling term shows little correlation with $n$, but a shallower $n$ means a lower normalization. These trends are reminiscent of the redshift trends we identified in Section~\ref{sec:results:nu}. To establish this connection more quantitatively, we compute an effective slope in \LCDM. The simplest definition is $\dnorminl{\ln(P)}{\ln(k)}$, but \citet{diemer_19_cm} showed that a tighter connection to concentration (and presumably to halo density profiles in general) is obtained by using
\begin{equation}
\neff(\nu, z) = -2 \left. \dnorm{\ln \sigma(R, z)}{\ln R} \right \vert_{R = \kappa R_{\rm L}}-3 \,,
\label{eq:neff-1}
\end{equation}
where $R_{\rm L}$ is the Lagrangian radius of a halo of a given peak height and redshift (Section~\ref{sec:methods:defs}) and $\kappa$ is a free parameter of order unity. Setting $\kappa = 1$ and $\nu = 1.75$ to conform with the peak height bin shown in the left column of \figmn{fig:av:cosmo}, we obtain $\neff = -1.78$ at $z = 0$ and $\neff = -2.4$ at $z = 4$. In the second column of \figmn{fig:av:cosmo}, we thus compare the \wmap results at $z = 0$ and $z = 4$ to the self-similar simulations with $n = -1.5$ and $n = -2.5$ (the former provides a closer match than $n = -2$). The correspondence is striking, both in the orbiting and infalling profiles and both in their normalization and slope. In the third and fourth columns of \figmn{fig:av:cosmo}, we additionally select haloes with $2 < \Gamma < 4$ to check that the trends with $n$ are not simply due to different mass accretion rates. Our previous conclusions largely hold for this sample, as well as for other $\Gamma$ bins. However, we observe that the truncation radius in \LCDM at $z = 0$ is smaller than that in the corresponding self-similar cosmology. A similar conclusion was reached by \citet{diemer_17_rsp}, who saw large splashback radii in self-similar universes with shallow $n$. 

Overall, the agreement between \LCDM and self-similar cosmologies demonstrates that the redshift evolution of profiles in \LCDM is better understood as a reflection of the changing power spectrum probed by haloes of different masses. Similarly, differences in the \LCDM parameters translate into different power spectra, whose slope affects the density profiles of haloes. The exact correspondence need not be simple though because the profiles can be affected by other haloes across a wide range of masses, e.g., due to satellite accretion. Definitions such as that of $\neff$ in \eqmn{eq:neff-1} collapse a complex dependence on the entire power spectrum into a single parameter.

\subsection{Halo-to-halo scatter in profiles}
\label{sec:results:scatter}

\begin{figure}
\centering
\includegraphics[trim =  5mm 8mm 162mm 2mm, clip, scale=0.65]{\figdir/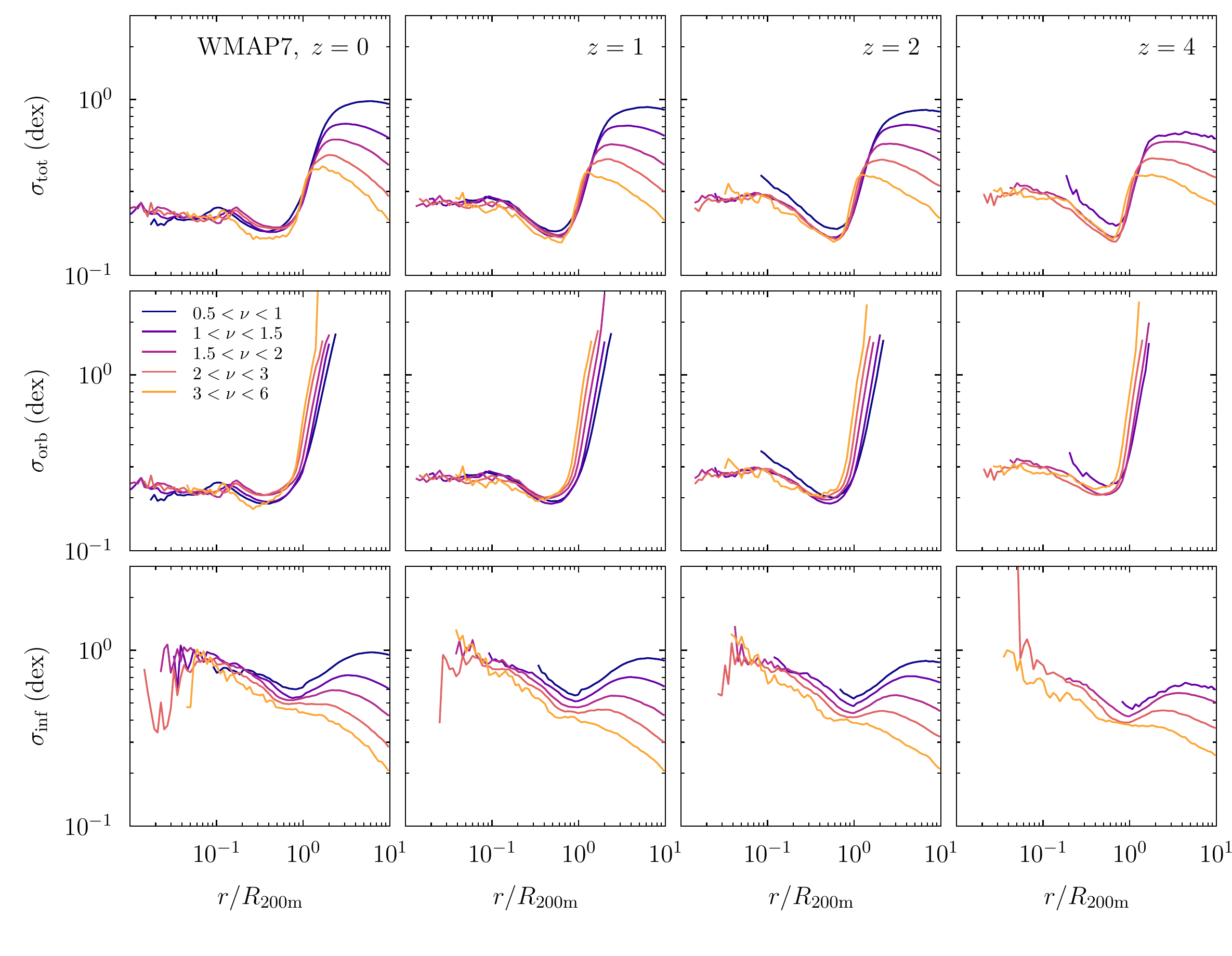}
\includegraphics[trim =  133mm 8mm 55mm 2mm, clip, scale=0.65]{\figdir/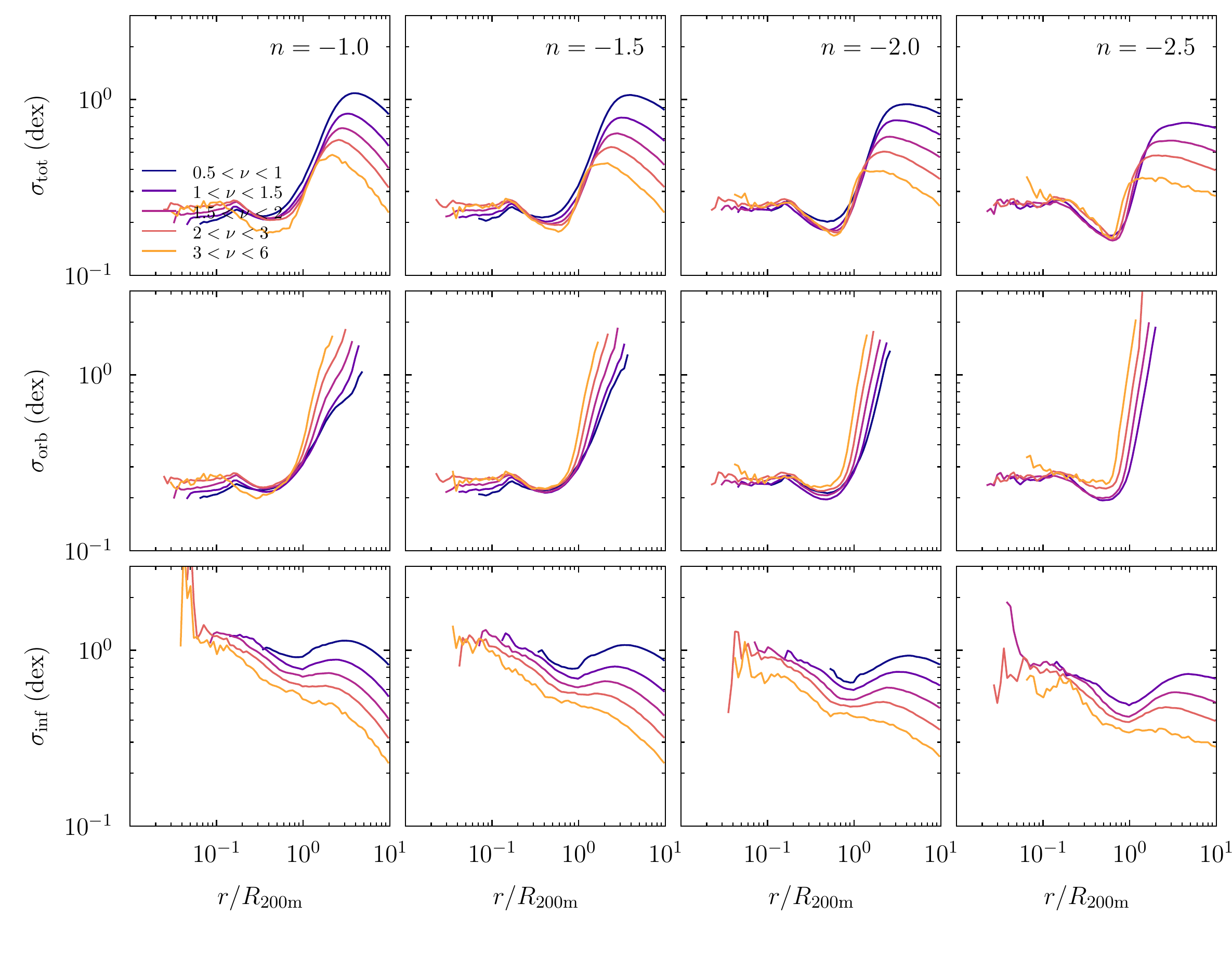}
\caption{Logarithmic scatter in dex around median profiles for mass-selected samples at $z = 0$ in the \wmap cosmology (left) and for the $n = -2$ self-similar cosmology (right). As in previous figures, the three rows show the scatter around the total, orbiting, and infalling profiles (from top to bottom). Plots for other redshifts and cosmologies appear very similar. The scatter in the total profiles is a strong function of radius: at $r \lsim \rtom$, it is more or less constant and takes on values roughly between $0.15$ and $0.35$ dex, regardless of mass, redshift, or cosmology. Around the truncation of the orbital term, the scatter increases dramatically due to different truncation radii in different haloes and due to different shapes of the edge of the orbiting term. The scatter in the infalling term is strongly mass-dependent because low-mass haloes are much more affected by their surroundings (see also \figmn{fig:prof:example}). Their scatter can reach $1$ dex at $r \gsim \rtom$, whereas the scatter around the most massive haloes remains below $0.5$ dex (although it increases towards the centre of haloes).}
\label{fig:scatter}
\end{figure}

We began the section by stating that individual halo profiles are widely scattered around their median, but how widely? \figmn{fig:prof:example} gives a visual example of the 68\% scatter around $\nu$-selected median profiles, showing a larger scatter in the infalling than orbiting profiles and in low-$\nu$ samples, which tend do be dominated by the environment rather than the haloes themselves. We confirm these impressions quantitatively in \figmn{fig:scatter}, which shows the radial dependence of the $68\%$ scatter in dex (that is, the logarithm of the ratio between the 84th to the 16th percentile). The scatter of the orbiting profiles is strikingly uniform across radii and peak heights, between $0.15$ and $0.35$ dex out to $r \approx 0.5\ \rtom$. Around the truncation radius, the scatter rapidly grows to arbitrary values as different haloes have slightly different truncation radii that lead to enormous fractional differences in density. Beyond the truncation radius, the scatter is dominated by that of the infalling profiles, which does depend on halo mass. While the highest $\nu$ bin experiences a scatter of about $0.2$ dex at $10\ \rtom$, the lowest bin experiences about $1$ dex. This picture is essentially independent of redshift and cosmology in \LCDM. Perhaps more surprisingly, the scatter profiles are also very similar in the self-similar simulations (right column of \figmn{fig:scatter}). When selecting by mass accretion rate in addition to mass, the scatter decreases as expected, but the rate of decrease depends on $\nu$ and $\Gamma$. For $\nu \gsim 0.5$ and $\Gamma \lsim 2$, the scatter falls to between $0.1$ and $0.2$ dex at small radii, whereas it remains similar to the purely mass-selected sample for the lowest $\nu$ bin. This trend suggests that low-mass haloes always suffer from larger scatter, even if selected by accretion rate. Moreover, the scatter gradually increases with $\Gamma$ and reaches $0.2$ to $0.4$ dex for $\Gamma > 4$. Some of this increase is caused by massive satellites, since high accretion rates often include large individual accretion events. The scatter in the outer profiles is only weakly affected by a $\Gamma$ selection. 

Overall, we conclude that the scatter in the inner profiles (inside the truncation radius) is modest, between $0.1$ and $0.4$ dex depending on the selection, mass, and accretion rate. The corresponding ranges are shown in \figmn{fig:prof:example} but would barely be visible in Figs.~\ref{fig:av:nu}, \ref{fig:av:nugamma:md}, and \ref{fig:av:cosmo}. On the other hand, the scatter in the infalling profiles is substantial, up to one dex for low-mass haloes \citep[see also][]{avilareese_99}.

%%%%%%%%%%%%%%%%%%%%%%%%%%%%%%%%%%%%%%%%%%%%%%%%%%%
% DISCUSSION
%%%%%%%%%%%%%%%%%%%%%%%%%%%%%%%%%%%%%%%%%%%%%%%%%%%

\section{Discussion}
\label{sec:discussion}

We have dynamically disentangled the orbiting and infalling components of halo density profiles and systematically investigated their shape. We have confirmed that the mass accretion rate of halos is the primary determinant of their profiles, although peak height, redshift, and cosmology also play a role (the latter two via the slope of the power spectrum). In this section, we discuss possible reasons for the physical origin of the profile shapes.

\subsection{The two-scale nature of the orbiting term}
\label{sec:discussion:two_scale}

On a particle level, the orbiting term represents the superposition of the phase space occupied by its constituent orbits \citep[e.g.,][]{lithwick_11, sugiura_20}, much like for stars in a galaxy \citep{schwarzschild_79}. Assuming that the distribution of total particle energies is limited by some maximum, there must be a corresponding radial cut-off roughly at the scale where the halo's potential reaches that maximum (as visualized in Figs.~\ref{fig:viz1} and \ref{fig:viz2}). Thus, the orbiting profiles exhibit two fundamental spatial scales. The first is some inner scale, whose existence has been observed ever since the early days of $N$-body simulations and which can be expressed as the scale radius $\rs$. The second scale is the truncation radius, which we shall denote $\rt$ following \citetalias{diemer_14}; it depends on the accretion rate (or, technically, its rescaled form $\rt / \rtom$ does). While the growth rate is correlated with concentration via the accretion histories of haloes \citep[e.g.][]{wechsler_02}, significant scatter in the $c$--$\Gamma$ and $\Gamma$--$\rt/\rtom$ relations means that haloes cannot generally be described by a single scale. For example, the truncation radii in \figmn{fig:av:nugamma:md} appear aligned due to the selection by $\Gamma$, but the profiles exhibit slightly different scale radii for different peak heights. 

The two-scale nature of the orbiting term has a profound consequence: density profiles cannot be `universal' in the sense that they are fully described by single-scale functions such as the NFW or Einasto profiles. While simulated profiles more or less follow those forms at $r \lsim \rtoc$, different dynamical states lead to different outer shapes. Conversely, lining up the profiles by their truncation radius cannot remove the dependence of the inner profile on concentration. This discussion begs the question of whether it makes sense to align the mean and median profiles at $\rtom$, rather than at $\rs$, $\rt$, or perhaps an entirely different radius \citep[e.g.,][]{garcia_21}. We would expect an alignment at $\rt$ to strongly reduce the scatter in the truncation region but to increase the scatter at smaller radii. We leave a detailed investigation to future work. In \papertwothree, we quantify $\rs$, $\rt$, and their relation based on a new fitting function.

In practice, one caveat to the two-scale nature is that not all haloes experience a sharp cut-off because some particles truly escape after orbiting, which can spatially extend the orbiting term to arbitrary distances. These extended profiles are much more prominent in low-mass haloes (\figmn{fig:av:nu}) and are mitigated by excluding haloes with a large unbound component (Appendix~\ref{sec:tests:neighbours}), indicating that the majority of escaping orbits are caused by close encounters with haloes of comparable size. Given that the vast majority of haloes do experience a sharp truncation, we leave a more detailed investigation of such unusual orbits for future work.

\subsection{Can we understand the shape of the truncation?}
\label{sec:discussion:truncation}

An intuitive theoretical explanation for the sharp truncation is provided by secondary infall models, where shells of dark matter are accreted onto an initial power-law overdensity \citep{fillmore_84, bertschinger_85, ryden_87, white_92}. These models predict an infinitely sharp caustic at the location of the last outermost orbiting shell (the splashback radius). While they naturally explain the general trend of this location with accretion rate \citep{adhikari_14, shi_16_rsp}, the agreement with simulations is only approximate \citep{diemer_17_rsp, sugiura_20}. Similarly, the smoother truncation in realistic profiles is readily explained by non-sphericity \citep[e.g.,][]{lithwick_11, vogelsberger_11_similarity, adhikari_14}, but there are currently no quantitative predictions as to what the realistic shape of the truncation should look like. The secondary infall model does, however, naturally predict the phase-space density profile of halos \citep{ludlow_10}, and it explains the second caustic in the slope in low-$\Gamma$ halos as particles on their second orbit \citep[\figmn{fig:av:nugamma:md},][]{adhikari_14}. The infinitely sharp shell in the model is transformed into a smoother, more realistic profile by effects such as non-sphericity \citep[e.g., fig.~3 in][]{ryden_93}. 

Alternatively, we might try to understand the shape of the truncation from fundamental physics. For example, similar cut-offs are seen in profile models that are based on assumptions about the distribution of particle energies \citep[e.g.,][]{king_66, hjorth_10_darkexp1}, but we show in \papertwo that the resulting models are not a good description of our orbiting profiles because energy-based cut-offs lead to infinitely steep profiles \citep{amorisco_22}. One reason for this disagreement might be that the halo particles do not really follow a single, simple distribution of energies \citep[e.g.,][]{pontzen_13}. Instead, their energies are determined by the collapse of a non-isotropic peak and by subsequently accreted subhaloes with their own energy distribution. Nonetheless, energy-based arguments can augment our understanding of mechanisms that cause a truncation of the phase-space distribution and thus of density profiles, for example in tidally stripped haloes \citep{drakos_17, amorisco_22}.

In summary, the sharp truncation is not a surprise given the results of shell models and energy-based calculations, but a detailed model of its shape will likely need to rely on a more detailed modelling of the distribution of particle orbits.

\subsection{Can we predict the slope of the orbiting term?}
\label{sec:discussion:orb_slope}

The secondary infall model also predicts the full density profile inside and outside the splashback radius, but it is well known that these predictions are not realistic, at least not unless the model is coupled with collapse from cored peaks \citep[e.g.,][]{dalal_10, ogiya_18, delos_19}, realistic accretion histories \citep{lu_06}, or mergers \citep{ogiya_16, angulo_17}. One reason is that, by itself, the model can predict only power-law density profiles because it evolves a self-similar initial density perturbation.\footnote{The secondary infall model does not necessarily make more accurate predictions for self-similar universes because a power-law power spectrum does not imply that the peaks are power laws in overdensity. As a result, the phase-space distribution of particles predicted by the model does not match simulated haloes even in self-similar cosmologies \citep{sugiura_20}.} While simulated profiles generally exhibit evolving slopes, approximate power-law shapes have been found under certain conditions \citep[e.g.,][]{tasitsiomi_04_clusterprof, anderhalden_13, delos_18}. We also find strikingly constant slopes in the regime of very rapidly collapsing haloes (Section~\ref{sec:results:nugamma}). In this section, we thus check whether the secondary infall model predicts the correct trends with accretion rate and power spectrum slope. 

We assume an initial overdensity with $\delta = \rho(<r) / \rhom \propto r^{-3 \varepsilon}$, where $\varepsilon \equiv -\dnorminl{\ln \delta}{\ln M(<r)}$. \citet{fillmore_84} showed that the resulting density profile is $\rho \propto r^{-\gamma}$ with a slope 
\begin{equation}
\label{eq:scm:gamma}
\gamma =  {\rm max} \left[ \frac{9 \varepsilon}{1 + 3 \varepsilon},\ 2 \right] \,.
\end{equation}
The perturbation cannot be steeper than $\delta \propto r^{-3}$ or $\varepsilon = 1$ because that corresponds to a point perturbation (the case treated in \citealt{bertschinger_85}), which gives $\gamma = 9/4$. The lower limit of $\gamma > 2$ can be understood as a consequence of the purely radial orbits assumed in the model: at radii much smaller than their apocentre, $r_{\rm apo}$, particles have approximately constant velocity and thus spend equal amounts of time in shells with volume proportional to $r^2$, leading to $\rho \propto r^{-2}$ \citep{dalal_10}. Thus, the secondary infall model produces power-law profiles with a relatively narrow range of slopes, $2 < \gamma < 2.25$. Comparing to the right columns of Figs.~\ref{fig:av:nugamma:md} and \ref{fig:av:cosmo}, we indeed find that the slopes of fast-accreting halos are close to $\gamma \approx 2$. The shallower inner profiles in simulations demonstrate the presence of tangential (or at least non-radial) orbits \citep[e.g.,][]{ryden_93, subramanian_00, lu_06}.

Regardless of the exact final slope, the secondary infall model predicts that the final profiles strongly depend on the initial conditions. The shape of the initial peaks is largely determined by the slope of the power spectrum \citep{bardeen_86}, and this shape imprints itself on the mass accretion histories and thus on the final profiles \citep{hoffman_85, cole_96, delpopolo_00, manrique_03, ricotti_03, ascasibar_04, lu_06, salvadorsole_07, dalal_10, ludlow_13, polisensky_15}. Mathematically, we can establish the connection between $n$ and $\gamma$ by assuming that the structure of a density peak is well described by the two-point correlation function, $\delta(r) \sim \xi(r)$ \citep[e.g.,][]{hoffman_85, ogiya_18}. In self-similar cosmologies, $\xi \propto r^{-3-n}$ and thus 
\begin{equation}
\label{eq:scm:n}
\varepsilon = -\frac{1}{3} \dnorm{\ln \delta}{\ln r} \sim -\frac{1}{3} \dnorm{\ln \xi}{\ln r} \sim 1 + \frac{n}{3} \quad \implies \quad \gamma \sim \frac{9 + 3n}{4 + n} \,,
\end{equation}
which would give $\gamma = 2 / 1.5 / 1$ for $n = -1/-2/-2.5$. This prediction is clearly invalid in the regime of purely radial orbits and depends on the model assumptions \citep[e.g.,][]{subramanian_00}. However, the general trend that higher $n$ lead to steeper profiles \citep[e.g.,][]{hoffman_85, lokas_00} is borne out in our comparison of self-similar cosmologies (\figmn{fig:av:cosmo}), although the range of slopes is much smaller than suggested by \eqmn{eq:scm:n}. The power spectrum slope has also been shown to affect the curvature of the orbiting term, where a steeper $n$ means a more rapidly steepening profile \citep{salvadorsole_12, ludlow_17, udrescu_19, brown_22_einasto}. We will quantify this effect parametrically in \paperthree.

We can also establish a connection between the predicted profile slope and the accretion rate. Typical structures collapse when the density inside their Lagrangian radius crosses the threshold $\deltac \approx 1.686$. By the definition of $\Gamma$, the Lagrangian mass is $M_\rmL \propto a^\Gamma \propto R_\rmL^3$ and thus $R_\rmL \propto a^{\Gamma / 3}$. The self-similarity of the model implies that $\Gamma$ is constant, meaning that it does not matter that we actually evaluate it over a dynamical time rather than instantaneously. Combining the previous expression with the linear evolution of overdensities in an $\Omega_\rmm = 1$ universe, $\delta \propto a$, and our definition of $\varepsilon$,
\begin{equation}
\delta \propto a R_\rmL^{-3\varepsilon} \approx \deltac \implies R_\rmL \propto a^{1/3\varepsilon} \implies \Gamma \sim 1/\varepsilon \implies \gamma \sim \frac{9}{\Gamma + 3}
\end{equation}
or, equivalently, $\dnorminl{\ln M}{\ln r} \sim 3 \Gamma / (3 + \Gamma)$ \citep{adhikari_14, shi_16_rsp}. For halos with $\Gamma = 0/1.5/3/6$ we predict $\gamma = 3/2/1.5/1$, or generally a shallower profile for higher accretion rates. This trend is borne out in \figmn{fig:av:nugamma:md}: all profiles reach $\gamma \approx 1.5$ at the smallest radii we can test, but $\Gamma \approx 0$ halos reach steepest slopes of $\gamma = 3$ at the truncation radius while $\Gamma > 2$ halos barely steepen past $\gamma = 2$. This small variation means that the profiles are, on average, fairly close to a power law, which reminds us of the argument for the asymptotic slope of $-2$: if a halo is accreting rapidly, many particles are on recently established orbits with apocentres near $\rt$, meaning that most of the density profile falls into the $r \ll r_{\rm apo}$ regime. Moreover, the orbits of freshly accreted particles have had less time to become isotropized, e.g., via the radial orbit instability \citep[e.g.,][]{merritt_85, macmillan_06}.

In summary, the secondary infall model can give insights into aspects of the formation of the orbiting profile and, with some additional assumptions, predicts the correct trends with $n$ and $\Gamma$ as well as power-law profiles with $\gamma = 2$ for fast-accreting halos. However, the detailed shape of the profile is a function of the entire mass accretion and merger history \citep{ludlow_13} and thus carries the complexities of non-linear structure formation. A theoretical model that fully describes the resulting profiles has yet to be formulated.

\subsection{What sets the shape of the infalling term?}
\label{sec:discussion:shape_inf}

The shape of the infalling term inside the transition region has hitherto been a mystery because its density is dominated by the orbiting term by many orders of magnitude. Our dynamical split has allowed us to systematically investigate the infalling term for the first time. Its exact shape at small radii could be fairly sensitive to the pericentre counting algorithm because the particle number per logarithmic radial bin declines towards the centre, leading to the omission of the infalling lines at radii smaller than $0.01$ to $0.1\ \rtom$ in Figs.~\ref{fig:prof:example}--\ref{fig:av:cosmo}. However, given our corrections for time binning effects (Section~\ref{sec:methods:prf}) and the results of convergence tests (Appendix~\ref{sec:tests}), we have no reason to suspect a systematic bias in the infalling profiles.

Overall, we find that the average infalling profiles are highly varied, with clear dependencies on mass, redshift, accretion rate, and cosmology (via the power spectrum). However, almost all infalling profiles steepen with radius and reach their steepest slope around the truncation radius before flattening at larger radii. The latter aspect depends on the plotted quantity though: the infalling profiles in overdensity, $\rho_{\rm inf} / \rhom - 1$, are fairly well fit by a single power law (\citetalias{diemer_14} and \papertwo). We have chosen to plot $\rho / \rhom$ because the overdensity can become negative. Regardless, it is clear that the infalling profiles do not share a common slope. 

What shapes might we expect for the infalling profiles? Perhaps the simplest expectation would be that infalling shells of dark matter contract gravitationally but do not cross, akin to applying the \citet{zeldovich_70} collapse model to spherical peaks \citep[e.g.,][]{mohammed_14}. This assumption leads to $\rho \propto r^{-3/2}$ \citep{bertschinger_85}, which is reflected in the overdensity slopes scattering roughly around $-1.5$ \citepalias{diemer_14}. We expect this picture to break down around the splashback (or truncation) radius, where shells cross for the first time. Indeed, we find that the profiles become shallower around this radius because infalling shells `feel' an interior mass distribution that decreases as they approach the halo centre.

Another simple, cosmology-dependent model for the outer profiles is to multiply a mass-dependent bias with the matter--matter correlation function, $\rhoinf / \rhom = b(\nu) \xi_{\rm mm}(r) + 1$ \citep[e.g.,][]{hayashi_08, tinker_10, tinker_12}. While the predicted profiles are relatively close to power laws, their slopes do not quite match simulations \citepalias{diemer_14}. Moreover, \figmn{fig:av:nu} makes it clear that there are significant slope variations with peak height, which cannot be captured by a scale-independent bias parameter. The predicted profile does evolve with redshift due to the evolving comoving scale probed by halos at fixed $\nu$ and due to the different time scalings of $\rhom(z)$ and $\xi_{\rm mm}(z)$, but the evolution does not match that of the simulated profiles. Nonetheless, the influence of cosmology on the outer profiles is strikingly borne out in our results, since the redshift evolution of the infalling profiles at fixed $\nu$ can be entirely explained by the effective slope of the power spectrum (Section~\ref{sec:results:cosmo}).

In summary, our results demonstrate that the infalling profiles are far more complex than simple power laws. Not only do they flatten at small radii, their slopes at large radii vary as a function of numerous halo properties and cosmology. We will further explore these dependencies in \papertwothree. 

%%%%%%%%%%%%%%%%%%%%%%%%%%%%%%%%%%%%%%%%%%%%%%%%%%%
% CONCLUSIONS
%%%%%%%%%%%%%%%%%%%%%%%%%%%%%%%%%%%%%%%%%%%%%%%%%%%

\section{Conclusions}
\label{sec:conclusion}

We have systematically investigated the density profiles of dark matter haloes in $N$-body simulations between $0.01$ and $10\ \rtom$, probing wide ranges of halo mass, redshift, and cosmology. For the first time, we have split the profiles into orbiting and infalling particles based on a novel algorithm that counts each particle's pericentres. Our main conclusions are as follows.
\begin{enumerate}

\item The orbiting profiles experience a sharp truncation at the edge of the orbital apocentre distribution. This splashback radius is generally apparent in the total profiles but not always located exactly at the radius where the total slope is steepest.

\item The truncation and scale radii exhibit different dependencies on halo properties, meaning that the orbiting profiles cannot be described as a function of a single scale parameter as suggested by many common fitting functions.

\item The mass accretion rate is the main factor determining the shape of density profiles. Slowly accreting haloes exhibit gradually steepening orbiting profiles whereas fast-accreting profiles tend towards power-law-like behaviour with mildly varying slopes. Peak height is only a secondary determinant of the orbiting profiles but has a significant impact on the infalling profiles.

\item The slope of the infalling term strongly varies from halo to halo, as well as with peak height and mass accretion rate, demonstrating that there is no uniform outer halo profile. 

\item The profile evolution with redshift in \LCDM can be understood as a variation in the effective slope of the power spectrum probed by haloes of different physical sizes, highlighting that density profiles are powerful diagnostics of cosmology. We can think of \LCDM as part of a continuum of self-similar universes with different power spectra.

\item The logarithmic 68\% scatter in the orbiting profiles is virtually constant inside the truncation radius and varies between $0.1$ and $0.4$ dex depending on the sample selection, peak height, and accretion rate. Lower mass haloes exhibit systematically larger scatter, especially in the infalling term where the scatter is mostly driven by the environment.

\item We demonstrate the convergence of our dynamical profile splitting algorithm with mass, force, and time resolution. However, we find that the density profiles are sensitive to definitions such as the halo boundary and the exclusion of merging systems.

\end{enumerate}
By splitting density profiles into their orbiting and infalling components, we can now systematically understand the profile shape in the transition region. However, numerous fundamental aspects remain to be explored, for example, an accurate theoretical model of the orbiting term based on particle dynamics. We have made our code and data publicly available in the hope of stimulating further investigations.

%%%%%%%%%%%%%%%%%%%%%%%%%%%%%%%%%%%%%%%%%%%%%%%%%%%
% ACKNOWLEDGMENTS
%%%%%%%%%%%%%%%%%%%%%%%%%%%%%%%%%%%%%%%%%%%%%%%%%%%

\section*{Acknowledgements}

I am deeply indebted to my PhD adviser, Andrey Kravtsov, for first suggesting halo density profiles as a research topic and for overriding my objection that everything worth knowing about them had already been explored. I thank Shaun Brown for feedback on a draft and the anonymous referee for their constructive suggestions. I am also grateful for enlightening discussions with Sten Delos (on theoretical collapse models) and with Phil Mansfield (on numerical convergence issues). Furthermore, I thank Susmita Adhikari, Han Aung, Rafael Garcia, Oliver Hahn, Daisuke Nagai, and Eduardo Rozo for productive conversations. This work was partially completed during the Coronavirus lockdown and would not have been possible without the essential workers who did not enjoy the privilege of working from the safety of their homes. The computations were run on the \textsc{Midway} computing cluster provided by the University of Chicago Research Computing Center and on the DeepThought2 cluster at the University of Maryland. This research extensively used the python packages \textsc{Numpy} \citep{code_numpy2}, \textsc{Scipy} \citep{code_scipy}, \textsc{Matplotlib} \citep{code_matplotlib}, and \colossus \citep{diemer_18_colossus}.

%%%%%%%%%%%%%%%%%%%%%%%%%%%%%%%%%%%%%%%%%%%%%%%%%%%
% DAVA AVAILABILITY
%%%%%%%%%%%%%%%%%%%%%%%%%%%%%%%%%%%%%%%%%%%%%%%%%%%

\section*{Data Availability}

The \sparta framework is publicly available in a BitBucket repository, \href{https://bitbucket.org/bdiemer/sparta}{bitbucket.org/bdiemer/sparta}. An extensive online documentation can be found at \href{https://bdiemer.bitbucket.io/sparta/}{bdiemer.bitbucket.io/sparta}. The \sparta output files (one file per simulation) are available in an hdf5 format at \href{http://erebos.astro.umd.edu/erebos/sparta}{erebos.astro.umd.edu/erebos/sparta}. A Python module to read these files is included in the \sparta code. Additional figures are provided online at \href{http://www.benediktdiemer.com/data/}{benediktdiemer.com/data}. The full particle data for the \erebos $N$-body simulations are too large to be permanently hosted online, but they are available upon request. 

%%%%%%%%%%%%%%%%%%%%%%%%%%%%%%%%%%%%%%%%%%%%%%%%%%%
% BIBLIOGRAPHY
%%%%%%%%%%%%%%%%%%%%%%%%%%%%%%%%%%%%%%%%%%%%%%%%%%%

\bibliographystyle{\includedir/citestyle_mnras}
\bibliography{\includedir/bib_mine.bib,\includedir/bib_general.bib,\includedir/bib_structure.bib,\includedir/bib_galaxies.bib,\includedir/bib_clusters.bib}

%%%%%%%%%%%%%%%%%%%%%%%%%%%%%%%%%%%%%%%%%%%%%%%%%%%
% APPENDIX: CONVERGENCE TESTS
%%%%%%%%%%%%%%%%%%%%%%%%%%%%%%%%%%%%%%%%%%%%%%%%%%%

\appendix

\section{Convergence tests}
\label{sec:tests}

\begin{figure*}
\centering
\includegraphics[trim =  2mm 9mm 0mm 3mm, clip, width=\textwidth]{\figdir/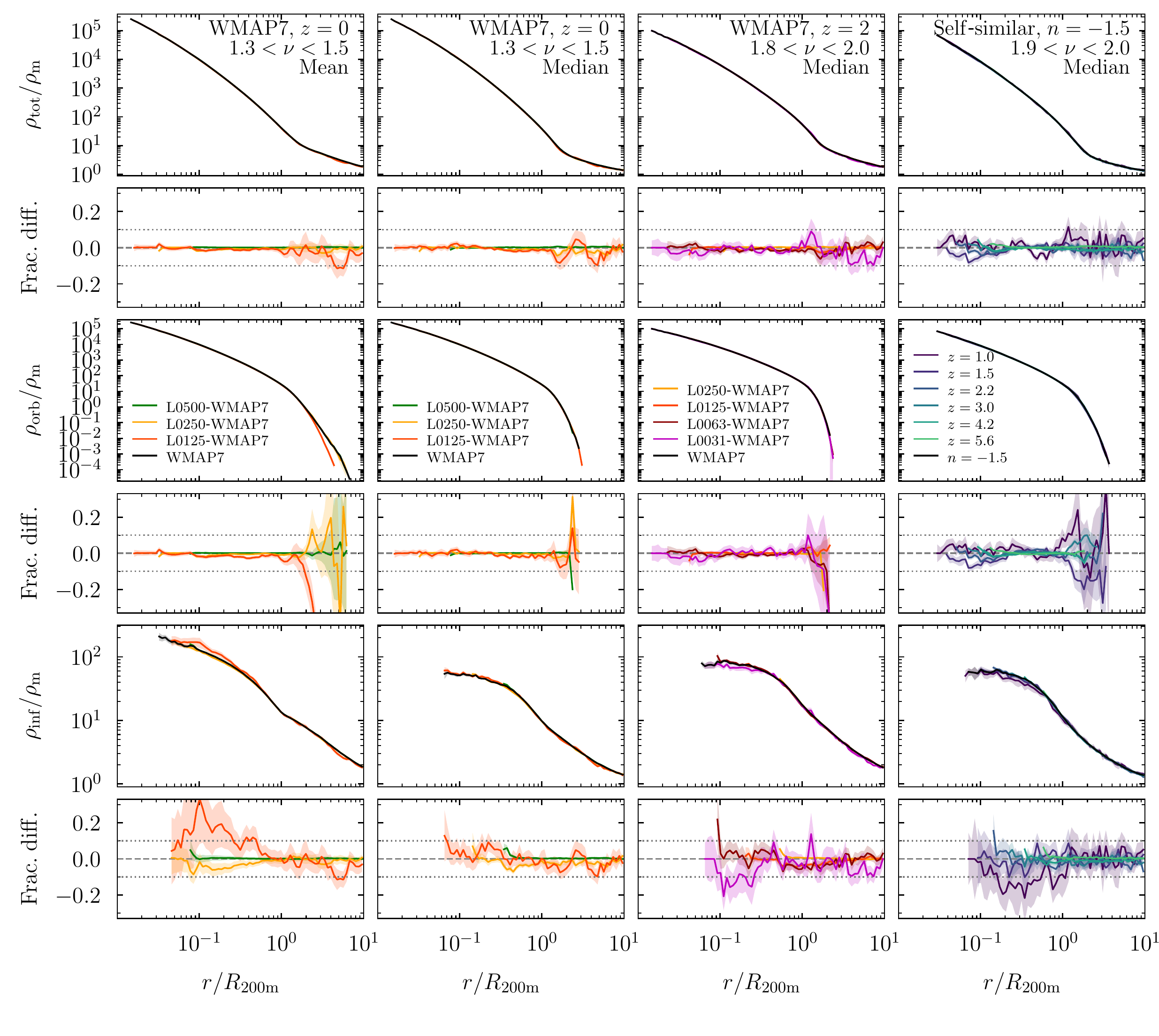}
\caption{Representative examples of convergence tests. Each set of panels shows density profiles (top) and the fractional difference (bottom) to the profile of the combined sample (black lines). The coloured lines refer to different simulation boxes of a \LCDM cosmology (left three columns) and to different redshifts in the $n=-1.5$ self-similar simulation (right column). Each set of boxes and redshifts corresponds to different resolutions for haloes at fixed peak height, and their agreement thus demonstrates convergence. As in the other figures, we do not show noisy profile bins with fewer than $80$ individual profiles.}
\label{fig:conv1}
\end{figure*}

\begin{figure*}
\centering
\includegraphics[trim =  2mm 9mm 0mm 3mm, clip, width=\textwidth]{\figdir/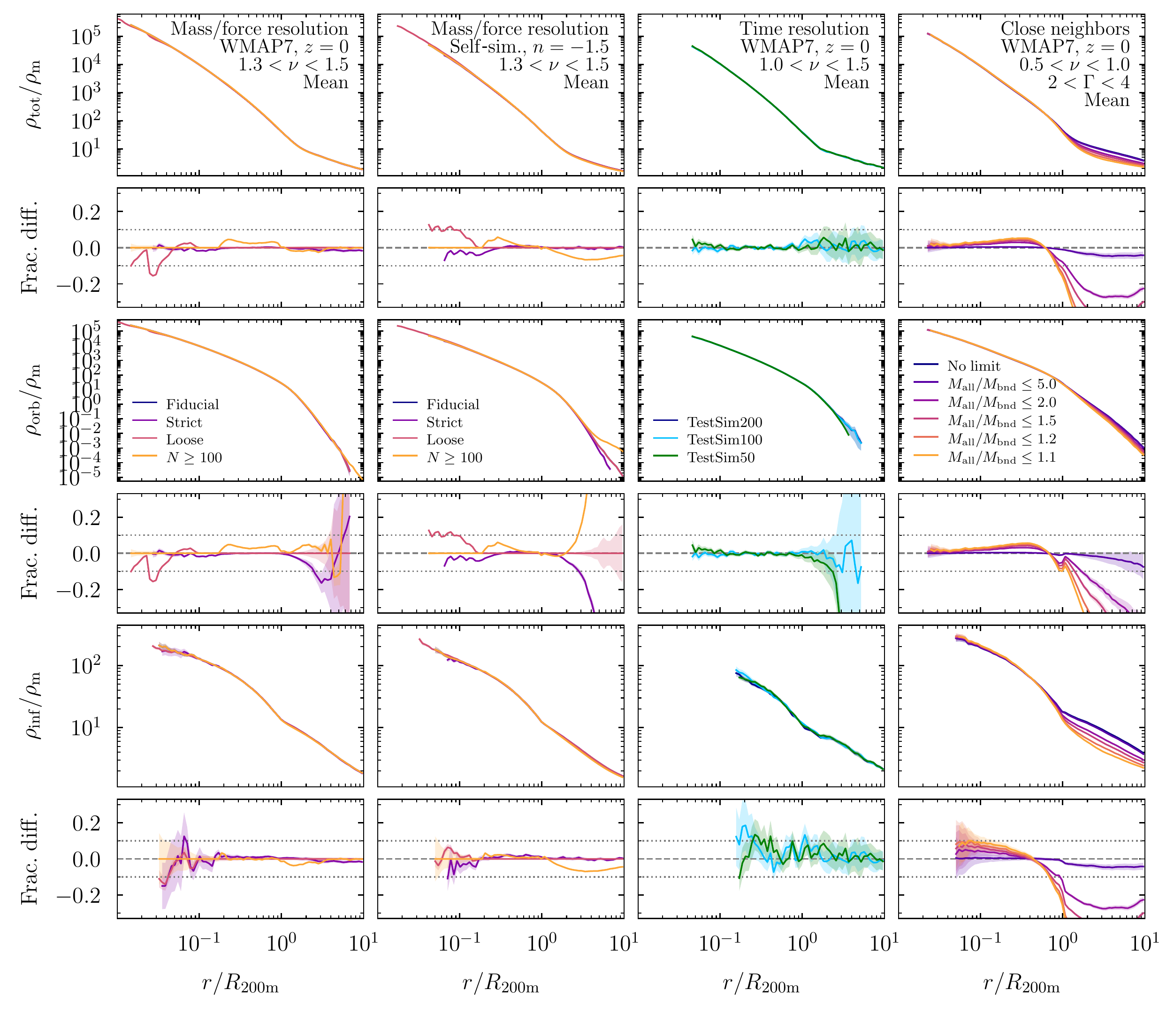}
\caption{Same as in \figmn{fig:conv1}, but instead of different simulation boxes we now compare samples defined by different resolution cuts (left two columns), time resolutions  (third column), and neighbour criteria (right column). In the left two columns, we observe no statistically significant differences beyond the expected 5\% convergence between our fiducial sample ($N \geq 500$, $r > 4 \epsilon$, $\kappa_{\rm P03} = 0.37$) and a version with stricter criteria (purple; $N \geq 2000$, $r > 8 \epsilon$, $\kappa_{\rm P03} = 0.74$), which indicates that our cuts in mass and force resolution are sufficient. This statement holds for both \LCDM (left column) and self-similar simulations (second column). The median profiles are even better converged than the mean. Samples with less stringent resolution criteria ($r > 1 \epsilon$ and $\kappa_{\rm P03} = 0.1$ in red, and $N \geq 100$ in yellow) lead to visible deviations. The third column compares profiles from our test simulation with three different time resolutions (snapshot spacings). We detect no significant differences between the high-resolution version (dark blue) and that with the usual $100$ snapshots (light blue). In the right column, we show the impact of our criterion to exclude contributions from nearby neighbours on low-mass haloes. As we decrease the allowed $\mtomall / \mtombnd$ ratio, the outer profile keeps decreasing and approaches the profile shape found for more massive haloes. Our chosen limit of $1.5$ represents a compromise.}
\label{fig:conv2}
\end{figure*}

In this appendix, we derive the resolution limits applied to our halo selection (\ref{sec:tests:res_limits}) and present tests that demonstrate the numerical convergence of our simulations and profile splitting algorithm (\ref{sec:tests:res_tests}--\ref{sec:tests:neighbours}). The figures shown are merely examples of a much larger range of data that we have inspected.

\subsection{Convergence limits}
\label{sec:tests:res_limits}

Any $N$-body simulation can produce reliable halo profiles only over a certain range of halo masses and spatial scales, and it is difficult to establish formal convergence criteria because different numerical effects can interact with each other \citep{moore_98, klypin_01, power_03}. Possible effects include spurious interactions between individual particles (known as two-body relaxation, \citealt{binney_08}), underestimated forces (and thus profiles) within a few force softening lengths of the centre of the halo, discreteness effects \citep{splinter_98, joyce_09}, and integration errors due to the finite time-steps. The latter two are thought to be subdominant, especially since the \textsc{Gadget} code uses adaptive timestepping that essentially ensures a certain number of time-steps per orbit at radii greater than the force resolution \citep{ludlow_19, mansfield_21_resolution}.

The remaining question is whether reduced centripetal forces or two-body relaxation are the more important, and how to limit their impact. As relaxation depends mostly on the particle number enclosed within a certain radius, the relative strength of the effects depends on $\epsilon / l$, the ratio of the force softening and inter-particle separation scales ($l \equiv L/N$). Our strategy is to compute the minimum radii where profiles should be converged with respect to either effect, and to use the more stringent of the two criteria. 

Specifically, we impose a target accuracy of $\sigma_{\rm res} = 5\%$ in the circular velocity profiles. The effect of force resolution was numerically investigated by \citet{mansfield_21_resolution} and encapsulated in the fitting function $\sigma_{\rm res}(r) = \exp [ -\left( Ah / r \right)^\beta]$ with $h = \epsilon / 0.357$, $A = 0.172$, and $\beta = -0.522$ for \textsc{Gadget} simulations. We invert this function and evaluate it for $\sigma_{\rm res} = 0.05$,
\begin{equation}
\label{eq:res:epsilon}
r_{\epsilon} / \epsilon = (A / 0.357) \left( - \ln \sigma_{\rm res} \right)^{-1/\beta} = 3.94 \approx 4 \,,
\end{equation}
where we have rounded to $r_{\epsilon} = 4\epsilon$ for convenience. To derive a comparable limit for the radius where we expect two-body relaxation effects to fall below 5\%, we rely on the \citet{power_03} parametrization of $\kappa_{\rm P03} \equiv t_{\rm relax} / t_{\rm orb}$, the ratio of the relaxation to orbiting time-scales. Analogous to \eqmn{eq:res:epsilon}, \citet{ludlow_19} provide a mapping between $\kappa_{\rm P03}$ and the accuracy of convergence (their equation~21), which we again invert to find
\begin{equation}
\label{eq:res:kappa}
\kappa_{\rm P03} = 10^{\psi - d} \,, \quad \psi = \frac{-b- \sqrt{b^2 - 4 a \left[ c - \log_{10} (1 - 10^{-\sigma_{\rm res}}) \right]}}{2a}
\end{equation}
with $(a, b, c, d) = (-0.4, -0.6, -0.55, 0.95)$. For $\sigma_{\rm res} = 0.05$, this formula gives $\kappa_{\rm P03} = 0.37$, meaning that the orbiting timescale should roughly be three times the relaxation time-scale in order to avoid resolution effects of more than 5\%. The radius where this criterion is satisfied, $r_{\rm conv}$, can be computed via an implicit relation that depends on the density profiles (and thus concentrations) of halos as well as on the simulation parameters \citep{power_03}. We approximate this calculation using equation~(18) of \citet{ludlow_19}, which is a function of only the inter-particle spacing, 
\begin{equation}
r_{\rm conv} = C \kappa_{\rm P03}^{2/3} \left( \frac{3 \Omega_{\rm m}}{800 \pi} \right)^{1/3} l(z) \,,
\end{equation}
where $C = 2.44$ and $l(z) = l / (1 + z)$ is the evolving inter-particle spacing in physical units. We have verified that this approximation provides a good description of the full, implicit calculation for our simulation parameters and the halo masses in question. Inserting our target accuracy of $\kappa_{\rm P03} = 0.37$, we find $r_{\rm conv} = 0.133\ \Omega_{\rm m}^{1/3}\ l(z)$, or $r_{\rm conv} = 0.086\ l(z)$ for the \wmap simulations, $r_{\rm conv} = 0.091\ l(z)$ for the \planck simulations, and $r_{\rm conv} = 0.133\ l(z)$ for the self-similar simulations.

For a given simulation, the minimum radii for both two-body relaxation and force resolution now correspond to a fixed comoving scale, meaning that we can directly compare their strength. For the \wmap simulations, two-body relaxation dominates over force resolution when $r_{\rm conv} > 4 \epsilon$ or $\epsilon / l < 1/47$ ($1/44$ in the \planck cosmology). Thus, two-body relaxation is the limiting factor in the $125\ \mpch$ and smaller boxes, and force resolution is the limiting factor in the $250\ \mpch$ and larger boxes. In the self-similar simulations, two-body relaxation dominates if  $\epsilon / l < 1/30$; given the very small softenings of $\epsilon / l < 1/195$ in the $n = -1$ and $n = -1.5$ simulations, the results are entirely limited by two-body relaxation in those boxes ($r_{\rm conv} \approx 7 \times 4 \epsilon$).

\subsection{Numerical convergence with mass and force resolution}
\label{sec:tests:res_tests}

We now numerically verify the criteria derived in the previous section by directly comparing haloes in different simulation at fixed peak height. Given that the simulations systematically vary in mass and force resolution, non-convergence would manifest as disagreements in profiles that should be indistinguishable. \figmn{fig:conv1} shows representative examples of our convergence tests, which include both the regimes where errors are dominated by force resolution (large boxes in \LCDM) and by two-body relaxation (small boxes and self-similar simulations). In the left three columns, we consider the \wmap cosmology at $z = 0$ and $z = 2$. The coloured lines show the mean (left column) and median (second and third columns) profiles from individual simulations as coloured lines. The black lines indicate the means and medians of the entire sample compiled from all simulations, to which we compare the individual simulations in the bottom panels. The shaded areas show our bootstrap uncertainties (Section~\ref{sec:methods:prf_av}). We restrict the profiles to our fiducial limits, namely, only halos with $N_{\rm min} \geq 500$ and only radial bins with $r_{\rm min} \geq {\rm max}(4 \epsilon, r_{\rm conv})$. Generally speaking, the total profiles are converged to within 5\% once the uncertainties are taken into account. This finding holds for both the mean and median profiles and at all redshifts.

The convergence of the total density profiles reflects our $N$-body simulations, but it does not imply that our algorithm to split particles into orbiting and infalling is also insensitive to resolution effects. We consider this more challenging test in the second and third rows of \figmn{fig:conv1}, which show that the split profiles are converged to roughly 10\% or better. The only notable exception occurs in the mean orbiting profiles at large radii, which fall to extremely low densities (e.g., $10^{-4} \rhom$ in the left column of \figmn{fig:conv1}). Here, individual outlier profiles can drastically affect the mean, for example, if a halo is tidally ripped apart. We should thus be cautious when interpreting the mean orbiting profiles at large radii.

For the self-similar simulations, convergence can be tested by comparing different redshifts of a single simulation. Given that both the power spectrum and time are independent of physical scales, the profiles at fixed peak height must match on average, except for resolution effects \citep{efstathiou_88, colombi_96, jain_98, widrow_09, joyce_21, leroy_21, maleubre_22}. The right column of \figmn{fig:conv1} shows profiles in a representative peak height bin for different redshifts in the $n = -1.5$ simulation. Once again, we find excellent convergence, even in the orbiting and infalling profiles. 

So far, we have compared halo samples that were selected with our fiducial resolution limits, $N_{\rm min} = 500$ and $r_{\rm min} \geq {\rm max}(4 \epsilon, r_{\rm conv})$. We visually justify these choices in the left two columns of \figmn{fig:conv2}, where we directly compare our fiducial sample to others with stricter and less strict cuts. We find no significant differences between the fiducial and stricter samples, which leads us to conclude that our limits are sufficiently strict. Furthermore, we have tested that the scatter around the mean and median values is not increased in our sample compared to higher resolution samples. On the other hand, less stringent cuts do have an impact on the mean. For example, including haloes with as few as $100$ particles leads to moderate, although statistically significant, deviations. Similarly, including those parts of profiles that suffer from poor mass and force resolution can lead to the characteristic down-turn of the profiles where they are underestimated due to a reduced forces and two-body scattering (left column of \figmn{fig:conv2} and Appendix~\ref{sec:tests:res_limits}).

Given that our profile splitting algorithm relies on particle orbits, it is subject to yet another potential resolution issue, namely the finite number of snapshots per orbit. For example, pericentres can be missed for orbits that are unresolved in time, which could lead to particles being erroneously classified as infalling (\figmn{fig:orbits}). We test the convergence with time resolution in the third column of \figmn{fig:conv2}. This test can only be performed with our smaller test simulation, where we can subsample $200$ snapshots. We find no significant differences between the full version and that subsampled to about $100$ snapshots at any mass or redshift. The slight differences in the total profiles (top row of \figmn{fig:conv2}) are due to the slightly different halo finder results. Even when subsampling by a factor of four, the differences are modest (green lines). These results give us confidence that the $100$ snapshots of the other \erebos simulations are sufficient (see \citetalias{diemer_17_sparta} for similar conclusions).

\subsection{Free parameters of the algorithm}
\label{sec:tests:parameters}

The pericentre detection algorithm presented in Section~\ref{sec:methods:oct} has a number of free parameters that could change the resulting split profiles. One such parameter is the radius at which particles are first tracked, which represents the largest radius at which pericentres can occur. This radius is set to $2\ \rtom$ (\citetalias{diemer_20_catalogs}). We find that increasing this radius does lead to a small number of additional pericentres and thus to a small shift of matter into the orbiting term. However, by visual inspection, we find that most of the additional pericentres appear to be spurious and due to irregular particle motions at large radii that cannot be clearly associated with a pericentre. Nevertheless, the uncertainty in how far out we allow pericentres slightly influences the shape of the orbiting profile at large radii, where the overdensity falls well below unity. Given that there simply is no `right answer' for what constitutes a pericentre, we conclude that these shapes inherently depend on our definitions to some extent.

Other parameters of the algorithm include the minimum angular distance a particle must have traversed at its first pericentre ($\phi < 0.5$) and the angular distance at which a pericentre is triggered in tangential orbits without clear radial velocity switches ($\phi = -0.8$). These values were determined by visual inspection of numerous orbits, but varying them within a reasonable range ($\phi \gsim 0$ as a minimum and $\phi \approx -1$ as a certain pericentre) does not noticeably change the profiles.

\subsection{Haloes with nearby neighbours}
\label{sec:tests:neighbours}

In Section~\ref{sec:methods:halo_sel}, we introduced a limit on the ratio of total and bound mass, $\mtomall / \mtombnd < 1.5$, to weed out haloes that are strongly affected by close neighbours. The impact of this cut on the mean profiles of low-mass haloes is shown in the right column of \figmn{fig:conv2}. Here, we consider haloes with high accretion rates where the trend is most visible, which is somewhat by construction as those haloes will typically have grown recently because they are `accreting' material from another halo. Cutting out such haloes reduces the density in the outer regions, which mostly affects the infalling profile but also has an effect on the orbiting term. The figure highlights that our choice of $\mtomall / \mtombnd < 1.5$ is a compromise between maintaining a complete halo sample and keeping the profiles free from the effects of nearby neighbours. Importantly, median profiles are much less affected than mean profiles because the effect is driven by relatively few extreme cases. Moreover, the convergence between simulation boxes (\figmn{fig:conv1}) is not affected by the neighbour criterion. 

Nearby haloes also appear to be at least partly responsible for the break in the slope that occurs around $\rtom$ in some infalling profiles. No such feature appears when haloes with nearby neighbours (or rather, even small fractions of unbound mass) are strictly excluded (yellow lines in \figmn{fig:conv1}), but it becomes gradually more prominent as the cut is relaxed. Similarly, the orbiting profiles of haloes with unbound material reach farther out, indicating that particles can be tidally removed by interactions. However, the exact change in the profile shapes due to neighbours is complex, and we defer a more thorough investigation to future work.

%%%%%%%%%%%%%%%%%%%%%%%%%%%%%%%%%%%%%%%%%%%%%%%%%%%%%%%%%%%%%%%%%%%%%%%%%%
% END
%%%%%%%%%%%%%%%%%%%%%%%%%%%%%%%%%%%%%%%%%%%%%%%%%%%%%%%%%%%%%%%%%%%%%%%%%%

\bsp
\label{lastpage}
\end{document}